%% file: main.tex
\documentclass[10pt,journal,compsoc]{IEEEtran}

\ifCLASSOPTIONcompsoc
  \usepackage[nocompress]{cite}
\else
  \usepackage{cite}
\fi

\usepackage{tabu}
\usepackage{float}
\usepackage{gensymb}
\usepackage{textcomp}
\usepackage{amsmath}
\usepackage{algpseudocode}
\usepackage{pifont}
\usepackage{verbatim}
\usepackage{float}
\usepackage{graphicx}
\usepackage{seqsplit}
\usepackage{caption}
\usepackage{enumitem}
\usepackage[hyphens]{url} 
\usepackage[hidelinks]{hyperref} \hypersetup{breaklinks=true}
\usepackage{booktabs}
\usepackage{mathtools}
\usepackage{multirow}
\usepackage{tabularx}
\usepackage{amssymb}
\usepackage{longtable}
\usepackage{ltxtable}
\usepackage{xtab,afterpage}
\usepackage{lipsum}
\usepackage[ruled]{algorithm2e}
\usepackage{subcaption}
\usepackage{pbox}
\usepackage{framed}

\usepackage{array}  
\usepackage[capitalise,noabbrev]{cleveref}

\usepackage{adjustbox}
\usepackage{diagbox}
\usepackage{threeparttable}

\usepackage{colortbl}
\usepackage{hhline}

\usepackage{tablefootnote}

\SetAlFnt{\small}
\SetAlCapFnt{\small}
\SetAlCapNameFnt{\small}
\SetAlCapHSkip{0pt}
\IncMargin{-\parindent}

\begin{document}

\title{Impact of E-Scooters on Pedestrian Safety: A Field Study Using Pedestrian Crowd-Sensing}

\author{Anindya Maiti$^\dagger$,~\IEEEmembership{Member,~IEEE,}
Nisha Vinayaga-Sureshkanth$^\dagger$,~\IEEEmembership{Student Member,~IEEE,}\\
Murtuza Jadliwala,~\IEEEmembership{Senior Member,~IEEE,}
Raveen Wijewickrama,~\IEEEmembership{Student Member,~IEEE,}\\
and Greg P. Griffin
\IEEEcompsocitemizethanks{
\IEEEcompsocthanksitem A. Maiti is with the University of Oklahoma, Norman, Oklahoma 73019, USA. \protect\\
E-mail: a.maiti@ieee.org
\IEEEcompsocthanksitem N. Vinayaga-Sureshkanth, M. Jadliwala, R. Wijewickrama, and G. P. Griffin are with the University of Texas at San Antonio, San Antonio, Texas 78249, USA. \protect\\
E-mail: vsnisha@ieee.org; murtuza.jadliwala@utsa.edu; raveen.wijewick@ieee.org; greg.griffin@utsa.edu
}
\thanks{$^\dagger$ These authors contributed equally to this work.}
\thanks{Research reported in this publication was supported by the United States National Science Foundation (NSF) under award number 1829066 (previously 1637290).}
}

\IEEEtitleabstractindextext{
\input{abstract}
}

\maketitle

\input{intro}

\input{background}
\input{goals}

\input{method}

\input{findings}
\input{discussion}
\input{conclusion}

\bibliographystyle{IEEEtran}
\bibliography{birdwatcher}

\input{appendix}

\vskip 0pt plus -1fil
\begin{IEEEbiography}[{\includegraphics[width=1in,height=1.25in,clip,keepaspectratio]{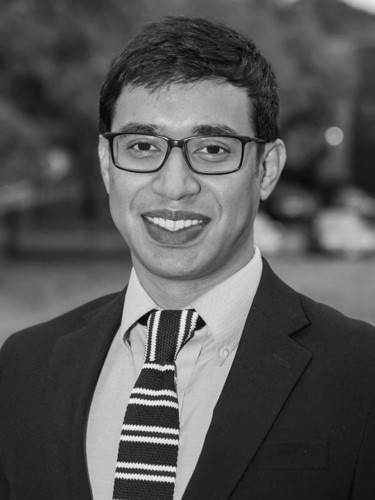}}]
{Anindya Maiti}
is currently an Assistant Professor in the Department of Computer Science at the University of Oklahoma, USA. Prior to that, he was a Postdoctoral Fellow in the Institute for Cyber Security at the University of Texas at San Antonio, USA, from 2018-2020. He received the Ph.D. degree in Electrical Engineering and Computer Science and the M.S. degree in Electrical Engineering from Wichita State University, USA, in 2018 and 2014, respectively. Prior to that he received the B.Tech. degree in Computer Science and Engineering from Vellore Institute of Technology, India, in 2012. His current research interests include vulnerability discovery and remediation in cyber-physical systems, and applied machine learning research in security and privacy.
\end{IEEEbiography}

\vskip 0pt plus -1fil
\begin{IEEEbiography}[{\includegraphics[width=1in,height=1.25in,clip,keepaspectratio]{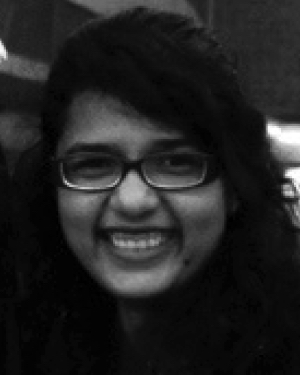}}]
{Nisha Vinayaga Sureshkanth}
is currently a Doctoral student at the University of Texas at San Antonio, USA. Prior to this, she obtained her Bachelor’s degree in Information Technology from Madras Institute of Technology, India. Her current research interests revolve around the areas of applied data science, and privacy and security in cyber-physical systems.
\end{IEEEbiography}

\vskip 0pt plus -1fil
\begin{IEEEbiography}[{\includegraphics[width=1in,height=1.25in,clip,keepaspectratio]{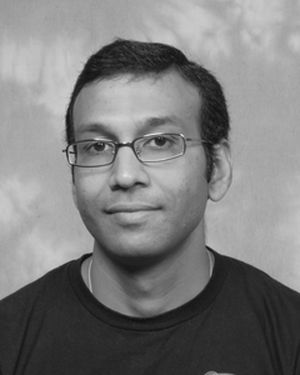}}]
{Murtuza Jadliwala}
is currently an Assistant Professor in the Department of Computer Science at the University of Texas at San Antonio, USA. Prior to that, he was an Assistant Professor in the Department of Electrical Engineering and Computer Science at the Wichita State University, USA from 2012-2017 and a Post-doctoral Research Fellow in the Department of Computer and Communication Sciences at the Swiss Federal Institute of Technology in Lausanne (EPFL) from 2008-2011. He also served as a Summer Faculty Fellow at the US Air Force Research Lab - Information Institute in Rome, NY, USA from June-August 2015. His educational background includes a Bachelors degree in Computer Engineering from Mumbai University, India and a Doctorate degree in Computer Science from the State University of New York at Buffalo, USA. His current research is focused towards overcoming security and privacy threats in networked computer and cyber-physical systems.
\end{IEEEbiography}

\vskip 0pt plus -1fil
\begin{IEEEbiography}[{\includegraphics[width=1in,height=1.25in,clip,keepaspectratio]{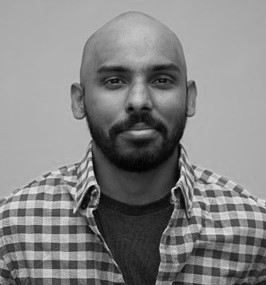}}]
{Raveen Wijewickrama}
is currently a Doctoral student at the University of Texas at San Antonio, USA. He received his M.S. degree from Wichita State University, USA in 2017 and his B.S. degree from Asia Pacific Institute of Information Technology (APIIT), Sri Lanka in 2015 in the field of Computer Science. His current research focuses on privacy and security in web services. His research interests also include privacy and security in wearable systems.
\end{IEEEbiography}

\vskip 0pt plus -1fil
\begin{IEEEbiography}[{\includegraphics[width=1in,height=1.25in,clip,keepaspectratio]{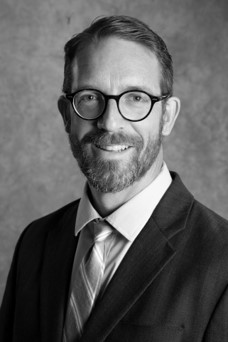}}]
{Greg P. Griffin}
is an Assistant Professor of Urban and Regional Planning, focusing on transportation, public participation, and health. His research involves how planners and publics work together with technology through three related topics of crowdsourcing, urban sensing, and co-production. He has over a decade of experience as a planner in Texas. A Dwight David Eisenhower Transportation Fellow, his doctoral program was in Community and Regional Planning at The University of Texas at Austin. He is certified by the American Institute of Certified Planners, continuously since 2005. His recent work is published in the Journal of the American Planning Association, Transportation Research Record, and in popular outlets including the Chicago Tribune.
\end{IEEEbiography}

\end{document}

%% file: abstract.tex
\begin{abstract}
The popularity and proliferation of electric scooters (e-scooters) as a micromobility solution in our cities and urban communities has been rapidly rising. Rent-by-the-minute pricing and a healthy competition between micromobility service providers is also benefiting riders with low trip costs. However, an unprepared urban infrastructure, combined with uncertain operation policies and poor regulation enforcement, has resulted in e-scooter riders encroaching public spaces meant for pedestrians, thus causing significant safety concerns both for themselves and the pedestrians. As a consequence, it has become critical to understand the current state of pedestrian safety in our urban communities vis-\`{a}-vis e-scooter services, identify factors that impact pedestrian safety due to such services, and determine how to support pedestrian safety going forward. Unfortunately, to date there have been no realistic, data-driven efforts within the research community that address these issues. In this work, we conduct a field study to empirically investigate %
crowd-sensed \emph{encounter} data between e-scooters and pedestrian participants on two urban university campuses over a three-month period. We also analyze encounter statistics and mobility trends that could identify potentially unsafe spatio-temporal zones for pedestrians. This first-of-its-kind work provides a preliminary blueprint on how crowd-sensed micromobility data can enable safety-related studies in urban communities.
 
\end{abstract}

%% file: intro.tex
\section{Introduction}

One of the biggest challenges faced by cities due to population growth and density is the transportation of commuters and intra-city travelers, especially over short non-walkable distances \cite{neves2019assessing}. A lack of adequate and/or frequent public-transportation infrastructure has partially catalyzed this situation in many cities \cite{pucher2009integrating}, which has resulted in increased use of personal automobiles, thus causing additional congestion on the roads. In addition to a sub-standard commute experience, this has also contributed to quality of life challenges, including an increase in air pollution levels with concomitant health and environmental problems \cite{kennedy2005four}, collisions \cite{crashstats-nhtsa} and economic waste \cite{cao2016built}. Due to these escalating problems with intra-city transportation, cities have deployed pilots and fully implemented systems of personal and service provider-owned \emph{electric} or \emph{battery-powered micromobility} vehicles.

\emph{Micromobility} is an umbrella term used to describe a novel category of transportation using non-conventional battery-powered vehicles aimed at shrinking the physical and environmental footprint required for quickly moving people over relatively short distances. Electric scooters (or e-scooters) \cite{sae} currently constitute the most popular class of micromobility vehicle \cite{scooter-investment}, which are designed for travel over distances that are too close to drive or utilize public transportation yet too far to walk \cite{shed}. Moreover, due to their small physical footprint, such vehicles provide a convenient means to navigate around a city with congested roads and sidewalks, thus making them a popular \emph{last-mile} transportation solution in urban areas \cite{mckenzie2019}. Last-mile transportation bridges the gap between conventional transportation hubs (such as a bus stop, train station and parking lot) and final destinations (such as a workplace, home, school, and shopping center), which is especially appealing in cities where conventional transportation options are not abundantly distributed. The popularity of e-scooters has been further accelerated by a growing number of service providers that offer these vehicles on \emph{rent-by-the-minute} schemes, wherein the riders do not have to bear the upfront purchase and maintenance costs of owning such vehicles. Other aspects of  e-scooter services that make them appealing to urban commuters include easy service accessibility through a smartphone application, flexibility in trip start and endpoints, ease of vehicle geo-location, the flexibility of drop-off options with no parking fees, a simplified and intuitive riding process which requires no pre-training and license to operate, and negligible environmental impact compared to fossil-fuel powered automobiles \cite{hardt2019usage}.

However, as with any disruptive new technology, unforeseen problems have surfaced with or due to such e-scooter services.
For instance, many city administrators and planners have been unable to cope with the sudden influx of e-scooters in their jurisdictions, and as a result, many urban jurisdictions have have had very lenient or no regulations on how these vehicles should be operated. 
As a result, e-scooter riders often end up encroaching road infrastructure meant for pedestrians, thus causing significant safety concerns both for themselves and the pedestrians \cite{fang2018riders}. Given that pedestrians face risks such as walking alongside riders traveling at high speeds and navigating around hazardously parked or standing vehicles on sidewalks, it is not surprising that a considerable number of reported micromobility vehicle incidents involve some form of collision with pedestrians \cite{sikka2019sharing}.
Furthermore, a study in Brisbane (Australia) found that nearly half of the shared e-scooter trips involved riding illegally in some way, such as riding on roads where it is not allowed, doubling with a passenger, or not wearing helmets (when required) \cite{haworth2019illegal}.

Thus, a critical issue that administrators, policymakers, and stakeholders in our urban communities need to address in a timely fashion is \emph{``how can pedestrians safely co-exist with e-scooters and e-scooters riders?''} As part of this overarching question, answers to specific questions such as ``\emph{what is the current state of pedestrian safety vis-\`{a}-vis e-scooter services in urban communities?}'', ``\emph{which factors impact pedestrian safety in such services?}'', and ``\emph{how to support pedestrian safety going forward?}'' are urgently needed. %
Public opinion both for and against such services has been highly polarizing which has resulted in abrupt responses from city administrators (e.g., some have welcomed e-scooters, while others have outright banned them \cite{anderson2019governing,riggs2020exploring,haworth2019illegal}) without clear justifications that are based on empirical data and analysis. \emph{Our position is that before making any policy decisions or implementing new regulations on e-scooter services, their impact on pedestrian safety needs to be thoroughly studied in an empirical and data-driven manner.}

Till date, there have been only a few research efforts that have attempted to empirically study the safety impacts of micromobility services in an urban environment \cite{oregonpilot-online,sikka2019sharing}. However, these efforts have primarily focused only on the problem of rider safety, either partially or wholly, leaving out the aspect of pedestrian safety impacted by these services. In this work, we conduct a field study to empirically investigate and characterize new safety issues that have arisen due to the introduction of e-scooter services, from the pedestrians' perspective. The most significant impediment in conducting such a field study is enabling pedestrians to collect and document information related to e-scooter movements and encounters, and its impact on their safety. An approach of asking pedestrians to document each and every encounter manually will be too cumbersome, error-prone, and exposed to bias. To overcome this challenge, we take advantage of the technical design of rental e-scooters, specifically, the onboard hardware and communication interfaces. Current service-provider owned e-scooters come equipped with a constantly beaconing Bluetooth Low Energy (BLE) radio, typically employed for near-field operations such as vehicle unlocking and communication with customers' mobile application. Our main idea is to passively capture the BLE signals/beacons continuously emitted by the BLE radios on-board these commercial e-scooters by employing pedestrian participants who are carrying some form of a BLE receiver (i.e., a smartphone or smartwatch). BLE and other sensor data crowd-sensed in such a fashion can then be used to extract fine-grained contextual (spatio-temporal) information about the mobility state(s) of the e-scooters and physiological states of the (participating) pedestrians. Using this information, and the resulting analysis, we will be able to better understand the various factors impacting pedestrian safety in such micromobility services.

Specifically, we conduct a field study by recruiting participants (mostly students) in the main and downtown campuses of the University of Texas at San Antonio, where e-scooter services are extremely popular. University campuses have a high density of pedestrians (who are also often distracted \cite{pedestrian-distraction}), making it an ideal environment for a field study such as this. %
We observed that it is not only possible to uniquely identify BLE beacons transmitted by e-scooters operated by popular service providers (e.g., Lime and Bird) on the above two campuses, but it is also possible to characterize encounters between these vehicles and pedestrians who are passively capturing these BLE beacons using their smartphones or smartwatches. Our field study focuses on crowd-sensing real-time e-scooter-pedestrian encounters and other pedestrian physiological data (such as heart rate) on the two campuses over three months by recruiting pedestrian participants and equipping them with customized BLE receivers such as smartwatches.

Well-defined spatio-temporal metrics are then computed from this crowd-sensed data and employed as safety benchmarks to further understand the impact that the e-scooter services operating on these campuses have on pedestrian safety. Our analysis uncovers interesting encounter statistics and mobility trends, which could be used to identify potentially unsafe spatio-temporal zones. %
Our study makes a preliminary effort to analyze the impact of new and upcoming micromobility transportation services on pedestrian safety and provides a blueprint on how relevant data crowd-sensed by pedestrians can be employed to conduct similar studies in other urban environments and communities.

%% file: background.tex
\section{Background and Related Work}
\label{bgr}
Before describing the research goals of this paper, we first present a brief background on e-scooter vehicles and services, followed by an outline of the related literature.

\subsection{Micromobility and E-scooters}
Several different types and form-factors of urban micromobility vehicles are being offered, primarily on a rent-by-the-minute rental model, by a range of service providers. Powered micromobility vehicle types include electric bicycles, boards, skates, and both seated and standing scooters \cite{sae}. Depending on the vehicle form-factor and target market, service providers may offer their vehicles in either a \emph{docked} or a \emph{dockless} model. In the docked model, vehicles may only be picked up and dropped off at specific locations, commonly known as docking stations. The dockless model offers more flexibility to riders as they can pick up and drop off the vehicles at any location within the geo-fenced area of operation. This model is relatively standard in small form-factor vehicles such as battery-powered e-scooters.

There are several reasons for focusing on dockless e-scooters in this study. First, e-scooters are currently the fastest growing form factor throughout the micromobility industry \cite{scooter-investment}. E-scooters services started in the United States in 2017, and quickly expanded to 110 cities by 2019 \cite{trend}. Second, any middle or large-sized city in the US is presently served by a large number of local and national e-scooter service providers, offering ubiquitously available vehicles and a range of different service options. Lastly, e-scooters are not only accessible for short-distance/last-mile trips within the city, but also for commuting within larger self-administered communities inside cities such as universities, schools, and company campuses and shopping malls. See \cref{tab:scooter-features} for the range of service providers and e-scooter types (and their features) found on our university campuses. 

{
\renewcommand\arraystretch{1.33}
\begin{table*}[]
\scriptsize
\centering
\caption{E-scooter service providers (in and around our university campus) and their vehicle features.}
\label{tab:scooter-features}
\begin{tabular}{c|c|c|c|c|c|c|c|c|c|c|c|c|}
\cline{2-8}
\multirow{2}{*}{} &
\multirow{2}{*}{Service Providers} &
\multicolumn{3}{c|}{Bird \cite{bird2019}} &
\multicolumn{2}{c|}{Lime \cite{lime2019}} &
\multicolumn{1}{c|}{Blue Duck \cite{blueduck2019}} \\

\cline{3-8} & &
Xiaomi M365 \phantom{.} &
\pbox{15cm}{\quad \\Segway Ninebot ES2 \\ (w/ extended battery)} \phantom{.} &
Electisan F350 \phantom{.} &
\pbox{15cm}{\quad \\Custom-made\\ Ninebot} \phantom{.} &
\pbox{15cm}{\quad \\Segway Ninebot ES2 \\ (w/ extended battery)} \phantom{.} &
Xiaomi M365 \phantom{.} \\

\hline
\multicolumn{1}{|c|}{\multirow{6}{*}{\rotatebox{90}{Features}}} &
Headlights & \checkmark & \checkmark & \checkmark & \checkmark & \checkmark & \checkmark \\
\cline{2-8} \multicolumn{1}{|l|}{} &
Tail/Brake Lights & \checkmark & \checkmark & \checkmark & \checkmark & \checkmark & \checkmark \\
\cline{2-8} \multicolumn{1}{|l|}{} &
Bell/Horn & \checkmark & \checkmark & \checkmark & \checkmark & \checkmark & \checkmark \\
\cline{2-8} \multicolumn{1}{|l|}{} &
Display & & \checkmark & \checkmark & & \checkmark & \\
\cline{2-8} \multicolumn{1}{|l|}{} &
Range ($mi$) & 18.6 & 15.5 (28.0) & 20-30 & 12-25 & 15.5 (28.0) & 18.6\\
\cline{2-8} \multicolumn{1}{|l|}{} &
Top Speed ($mph$) & 15.5 & 15.5 (18.6) & 18.0 & 15.5 & 15.5 (18.6) & 15.5 \\
\hline
\end{tabular}
\end{table*}
}

Although e-scooters are available for a personal purchase, %
it is the \emph{servitization} of these vehicles that have resulted in their popularity. Servitization allows riders to use the nearest available vehicle, which should be easy to find in an urban setting due to a large density of vehicles, without having to securely store or carry them along when not in use. Vehicle rental (pick-up and drop-off), vehicle geo-location, service tracking, and payments are facilitated through mobile apps implemented by the service provider. In addition to the on-demand nature of these services, the offered vehicles are environmentally friendly when micromobility trips replace personal automobile use \cite{hollingsworth2019scooters}.

Renting and operating these vehicles is fairly straightforward. Through the service provider's smartphone application, riders can activate any available e-scooter belonging to the provider that they find nearby and pay to ride it for as long as needed, or until the battery is drained. Riders can travel up to 28 miles per charge on certain e-scooter models, but most e-scooter trips are typically much shorter \cite{hardt2019usage}. These vehicles fit the SAE ``low-speed'' category, with top speeds less than 20 mph \cite{sae}. Riders typically pay anywhere between 15 to 50 cents per minute to use the e-scooters, but some service providers also charge a base fee (currently \$1.00 for Bird and Lime) to activate an e-scooter. The overall cost is significantly lower compared to minimum fares of popular automobile ride-hailing services such as Uber and Lyft (approximately \$8.00 in the United States, with slight variation between cities). Riders are also expected to educate themselves and comply with the local laws and regulations, e.g., wearing a helmet and riding only in bike lanes, while riding these vehicles.

\subsection{Pedestrian Safety in the Built Environment}
Any vehicle sharing space with pedestrians poses a risk of collision, but the threat can vary due to urban density, pedestrian infrastructure, roadway design, traffic volume and speed, visibility, and the type of pedestrian \cite{Stoker2015}. Philip Stoker et al.'s systematic review of over 170 pedestrian safety studies showed the three primary factors that may mitigate risk are also largely controllable through planning and design: pedestrian-traffic interaction, visibility, and traffic speed \cite{Stoker2015}. A comprehensive review of street design factors showed streets with sidewalks on both sides, slower traffic speeds, buffers and barriers, landscaping, and trees all supported reduced pedestrian risk, in addition to 20 other factors \cite{Asadi-Shekari2015}. However, the ways that these variables impact safety are uneven across communities.

Low-income and minority communities experience greater risk while walking, as compared with higher-income and White populations. Observations of drivers at crosswalks show that they yield to Black pedestrians at half the rate, with wait times 32\% higher than White pedestrians \cite{Goddard2015}. Research on disparities in pedestrian risk shows evidence for prioritizing safety improvements in areas with high rates of minorities and poverty, particularly near schools \cite{Yu2018}.

University campuses and surrounding areas pose a risk for pedestrians, including small vehicles on pathways, and with motor vehicles on campus fringes. Pedestrians perceive risk from bicycling in campus settings, showing importance in the travel experience, even when the issue is more of comfort than safety \cite{Gkekas2020}. Police statistics under-represent the number of pedestrian and bicycle crashes, supporting a role for crowdsourcing incident data \cite{Medury2017}. We find no research till date detailing differences in pedestrians' perceptions of e-scooters versus traditional bicycle modes.

\subsection{Prior Work on Safety Issues due to Micromobility Vehicles}
Prior research efforts to identify and/or address issues related to micromobility, especially regarding the safety of pedestrians and riders, did not have a holistic view of the underlying pedestrian and rider movement patterns. Analysis by micromobility service providers \cite{BirdSafe59:online}, who can easily gather contextual data related to their vehicles (such as riding patterns and parking habits), did not have any quantitative information on fellow pedestrians and their movement patterns. Moreover, service providers would have a business incentive to not highlight the negative impacts on pedestrian safety due to their vehicles. Similarly, studies by some city governments and community administrators \cite{oregonpilot-online,ontgomeryparks:online} only employed subjective feedback and qualitative data (often, more from pedestrians than riders).

Independent research efforts on micromobility related issues have thus far been very limited in scope. Initial studies took a broad approach to apply planning lessons from similar modes, and identify research needs \cite{cohen2016planning,berge2019kickstarting}. An observational study in west Los Angeles identified safety risks related to e-scooter driver behaviors, such as the ability to move between sidewalks and motor vehicle lanes, which may surprise motorists \cite{todd2019behavior}. In Singapore, researchers measured improvements in rider predictability after installation of directional arrows on paths, suggesting opportunities to improve safety through engineering for emerging modes \cite{lim2019research}. An early field study in China observed e-scooter riders to more often ride against the flow of traffic and in motorized lanes \cite{bai2015comparative}. Researchers from medical institutions have analyzed micromobility related injuries of both riders and pedestrians \cite{trivedi2019injuries,badeau2019emergency,aizpuru2019motorized,siman2017casualties}, and found that musculoskeletal fractures and head injuries were most common. While riders may be compelled to wear proper protective gear based on these findings (for example, mandatory use of helmets as suggested by Choron and Sakran \cite{choron2019integration}), the same cannot be enforced on fellow pedestrians. Sikka et al. highlighted the health and financial impact for pedestrians involved in an e-scooter collision, using a case study \cite{sikka2019sharing}.

New approaches connect safety research to mobility needs, leveraging observational data. McKenzie analyzed usage patterns of e-scooter and e-bikes in Washington, DC, using the city's publicly accessible API to micromobility data portals \cite{mckenzie2019}. James et al. analyzed e-scooter safety perceptions and sidewalk blocking frequencies from survey data, and observed parking practices in different built environments \cite{james2019pedestrians}. The use of virtual reality enables controlled experimentation of different e-scooter safety contexts without risk of field interventions \cite{gossling2020integrating}. Initial empirical results support additional policy-focused work to integrate micromobility as part of a sustainable transportation system \cite{che2020users}. Micromobility research increasingly leverages new data collection methods to address a wide range of needs. Yet, none to date evaluate a system for pedestrian-focused e-scooter interaction.

In this work, we systematically analyze e-scooter and pedestrian encounters (a precondition to accidents involving e-scooters and pedestrians), and discern if or how pedestrians and such micromobility services can safely co-exist in urban environments.

%% file: goals.tex
\section{Research Objectives}
\label{goals}

Disruptions to pedestrian movement due to micromobility vehicles such as e-scooters, and collisions between these vehicles and pedestrians, occur only when they closely (in some spatio-temporal sense) \emph{encounter} each other on the streets. A more precise and empirically derived definition of an encounter is detailed later in \cref{encounter-definition}. 
Given the significant number of incidents involving pedestrians and micromobility vehicles reported in the last two years \cite{sikka2019sharing}%
, we can postulate that every such close encounter between micromobility vehicles (moving or stationary) and pedestrians has some probability of resulting in a collision or a disruption to pedestrian movement. In other words, a higher density/concentration or frequency (or both) of such close vehicle-pedestrian encounters is indicative of a higher probability or potential for vehicle-pedestrian collisions and is generally a good metric for benchmarking the state of pedestrians' safety.

There are two critical factors that dictate the occurrence of close vehicle-pedestrian encounters, and their density and frequency. The first, also referred to as \emph{space factors}, are the \emph{spatial constraints} imposed by the infrastructure (roads, sidewalks, etc.) shared by the micromobility vehicles and pedestrians. The second, also referred to as \emph{time factors}, are the \emph{temporal constraints} that dictate the mobility (speed, direction, etc.) of the micromobility vehicles and pedestrians within a given shared space. A \textit{combination or co-existence} of these space and time factors also impact the occurrence of encounters. In order to further clarify this, let us give some concrete examples of these factors as observed by us during our study. 

\begin{figure}[]
\centering
\begin{subfigure}{0.49\linewidth}
\centering
\includegraphics[width=\linewidth]{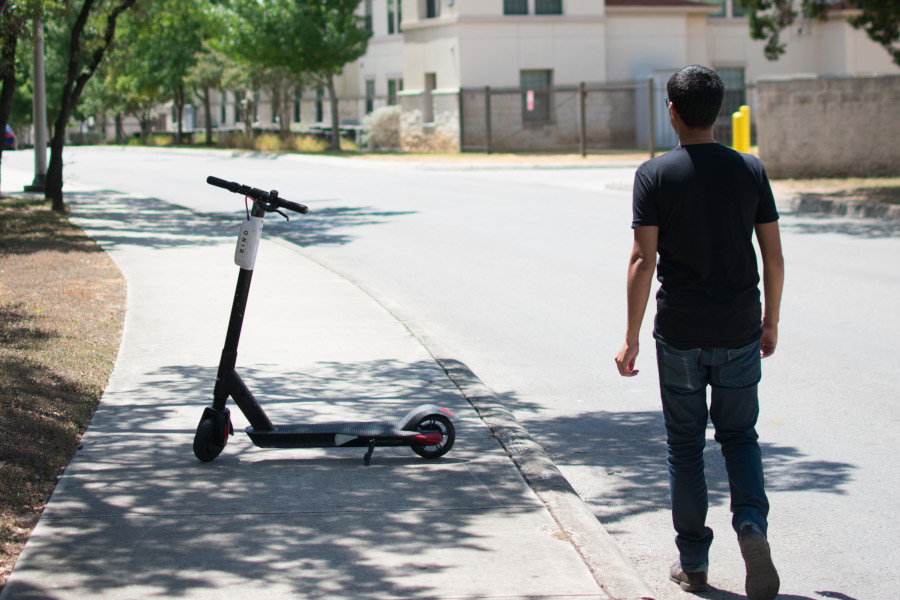}
\captionof{figure}{\scriptsize Improperly parked e-scooter.}%
\label{fig:sidewalk-bypass}
\end{subfigure}
\begin{subfigure}{0.49\linewidth}
\centering
\includegraphics[width=\linewidth]{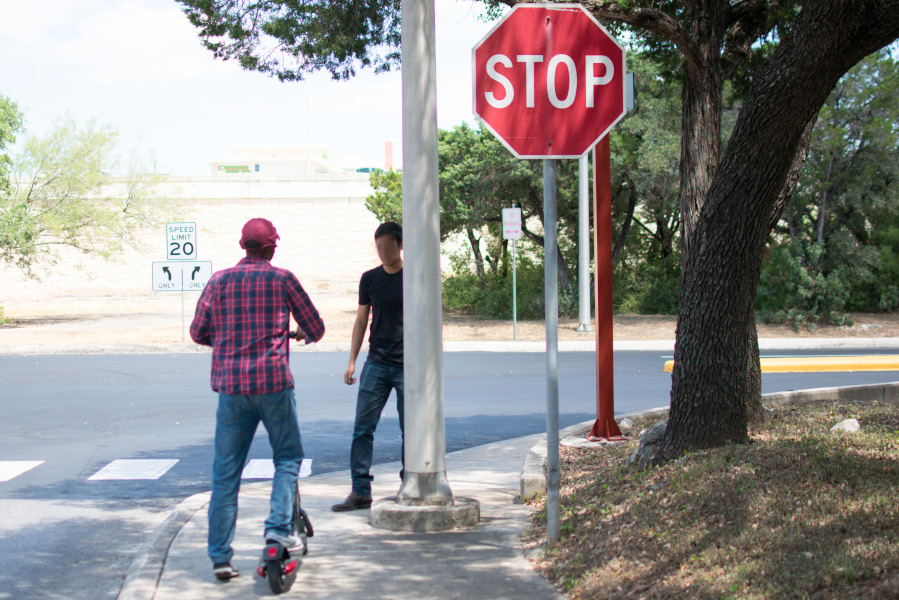}
\captionof{figure}{\scriptsize A street light pole.}%
\label{fig:sidewalk-obstruction}
\end{subfigure}
\caption{Scenarios with pedestrian path roadblocks.}
\end{figure}

For instance, insufficient allocation of space for sidewalks and bike lanes can lead to unsafe encounters between e-scooters and pedestrians. If a bike lane is not present, e-scooter riders may feel compelled to use sidewalks meant for pedestrians. Similarly, if an improperly parked e-scooter is blocking a sidewalk, pedestrians may be forced to use the main road to bypass the blockade (as shown in \cref{fig:sidewalk-bypass}) which places them in great danger of getting hit by cars on the road. Other permanent obstructions, for example, trees, poles (as shown in \cref{fig:sidewalk-obstruction}), benches and fire hydrants, on spaces often shared between e-scooter riders and pedestrians can also lead to unsafe encounters. Safe utilization of space allocated to riders and pedestrians also depends on proper planning of transportation hubs. For instance, if all commuters who just got off a bus head in the same direction to their final destination, it may cause congestion among riders and pedestrians covering their last-mile. An optimally positioned bus stop, train station or parking lot should observe the diffusion of commuters in all directions, thus minimizing chances of congestion and making safer utilization of the space allocated for riders and pedestrians.

Similarly, several time factors also play an important role in generating potentially unsafe micromobility vehicle-pedestrian encounters within a given space. For instance, if there is a spike in rider and pedestrian traffic due to multiple closely timed events (e.g., multiple classes scheduled in the same building and starting at the same time), it may cause congestion among riders and pedestrians en-route to these events. Another crucial time factor is the reaction time pedestrians get to navigate around micromobility riders traveling at different speeds and in different directions. Depending on whether a micromobility vehicle is moving towards or away from a pedestrian, and whether the vehicle is behind or in front of the pedestrian, the pedestrian may or may not get sufficient time to react appropriately.

Our research agenda, thus, is to first analyze by means of empirically collected encounter data how certain \emph{space} and \emph{time} factors affect the safety state of pedestrians when they are in co-existence with e-scooters (and riders). Specifically, we seek to conduct the following three broad research analyses:

\begin{enumerate}[label=RA\arabic*, leftmargin=2.5em]
\setcounter{enumi}{0}
\begin{framed}
\item
\emph{Correlating \textbf{space factors} with empirical encounter and physiological data to identify potentially unsafe (to pedestrians) encounters and contexts.} \label{rq:space}
\end{framed}
\end{enumerate}

Specifically, in \ref{rq:space}, we analyze the spatial distribution of encounters, changes in encounter properties between high and low encounter concentration or density areas, and the effects of pedestrians' and riders' spatial diffusion on encounter rates and other encounter-related properties in order to understand their impact on pedestrian safety. We will also relate this analysis to infrastructure-related shortcomings, such as missing bike lanes and sidewalk obstructions, in order to determine potentially unsafe encounters, if any. %

\begin{enumerate}[label=RA\arabic*, leftmargin=2.5em]
\setcounter{enumi}{1}
\begin{framed}
\item
\emph{Correlating \textbf{time factors} with empirical encounter and physiological data to identify potentially unsafe (to pedestrians) encounters and contexts.} \label{rq:time}
\end{framed}
\end{enumerate}

Specifically, in \ref{rq:time}, we analyze the temporal distribution of encounters, changes in encounter properties between time periods comprising of a large number of encounters versus the smaller number of encounters, and the effects of pedestrians' and riders' temporal diffusion on encounter rates and other encounter-related properties in order to understand their impact on pedestrian safety. 
As before, we will relate this analysis to the infrastructure-related shortcomings, such as unbalanced class schedules and common event times, in order to determine potentially unsafe encounters, if any. Additionally, we will also analyze different encounter scenarios that give varying levels of reaction time to pedestrians and quantitatively measure pedestrians' reactions to these different encounter scenarios. 

\begin{enumerate}[label=RA\arabic*, leftmargin=2.5em]
\setcounter{enumi}{2}
\begin{framed}
\item
\emph{Correlating a combination of \textbf{space} \& \textbf{time factors} with empirical encounter and physiological data to identify potentially unsafe (to pedestrians) encounters and contexts.} \label{rq:space-time}
\end{framed}
\end{enumerate}

In \ref{rq:space-time}, we will extend our previous analyses to study which combinations of space factors (e.g., poor shared space utilization) and time factors (e.g., event times), as discussed earlier, are the most significant enablers of unsafe encounters between pedestrians and riders. 

In addition to the above quantitative analyses, which are primarily based on the crowd-sensed (BLE-based) encounter data and data from mobile sensors (e.g., heart rate), we will also analyze pedestrians' attitude and perception towards the impact that e-scooters have on pedestrian safety (\cref{subsec:perceptions}).

%% file: method.tex
\section{Research Methodology}
\label{mtd}

We now describe the details of the field study that we conducted for crowd-sensing the e-scooter--pedestrian encounter and other pedestrian-specific data used for the safety analyses summarized earlier. As part of this description, we outline in detail the study environment, data collection process, including participant recruitment and the type and granularity of the data that is collected.

\subsection{Significance of Pedestrian's Point of View}
Let us first briefly describe why pedestrians are best suited for gathering (and crowd-sensing) detailed information on their encounters with e-scooters. The e-scooters may or may not have a rider at the time of an encounter (for example, a parked vehicle), which means we will fail to gather information on encounters between pedestrians and rider-less vehicles if we depend only on riders for data collection. The vehicles themselves feature several sensing options, but, (i) none of the on-board sensors are suitable for detecting nearby pedestrians, and (ii) service providers are not comfortable with releasing their vehicles' data due to potential misuse by competitors and customer/rider privacy concerns.

\begin{figure}[b]
\centering
\includegraphics[width=\linewidth]{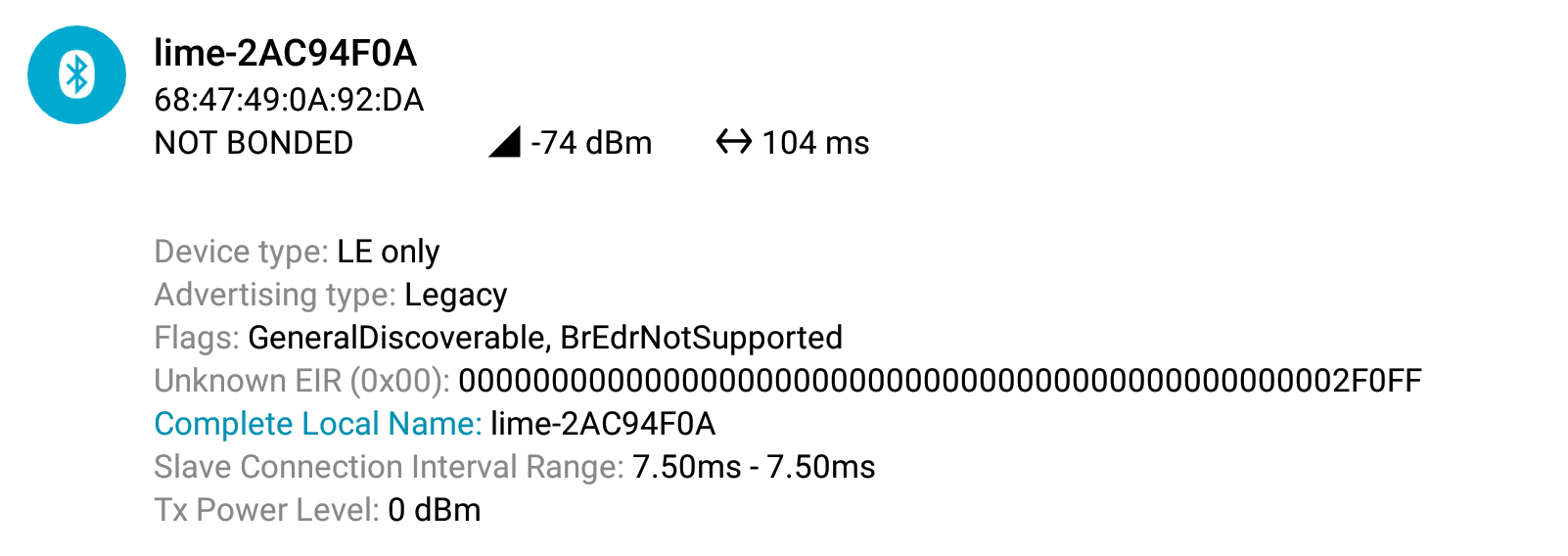}
\captionof{figure}{A BLE advertising packet from a Lime e-scooter.}
\label{fig:ble-example}
\centering
\end{figure}

Pedestrians also carry a variety of sensors with them that are present on their mobile and/or wearable devices. While experimenting with different sensors that could be employed for detecting encounters, we determined that most e-scooters transmit BLE advertising packets at regular intervals, which could be passively captured by the BLE receivers present on most smartphones or wearables carried by the pedestrians. These BLE packets also contain identifiers that can be used to distinguish them from other BLE devices. For example, they may contain the service provider's name (as shown in \cref{fig:ble-example}) or other unique naming conventions. Furthermore, due to the short range of BLE transmissions, pedestrians may capture the BLE packets only when they encounter a nearby e-scooter, which can minimize unwanted noise that could occur due to e-scooters that are not close to the pedestrian which also reduces the task load (\cref{data-collection}) for our participants. Further, it also obviates participants having to carry any specialized sensing hardware, and it is reasonable to assume that most pedestrians are comfortable and used to carrying a smartphone or wearable such as a smartwatch.

\subsection{Data Collection} \label{data-collection}
In order to accomplish the research goals outlined in \cref{goals}, we crowd-sensed real-life e-scooter--pedestrian encounter data by capturing BLE packets emanating from e-scooters in two separate urban communities supplemented by physiological and contextual (location and time) information and real-time feedback from the participating pedestrians. 

\noindent
\textbf{The Field:}
To have a controlled understanding of encounters, we limited the field of our study to the main and downtown campuses of the University of Texas at San Antonio and neighboring points-of-interest including off-campus student housing and transportation hubs. Both campuses are within the city perimeters and cover about 725 acres in total area. As an urban university with more than 35,000 students and more than 4,000 employees, our campuses observe significant foot traffic when classes are in session. Since their introduction in late 2018, e-scooters have gained significant popularity throughout the city, including our university campuses. Students and employees primarily use micromobility services as a last-mile solution on campus, e.g., to travel between parking lots, bus stops or student housings, and university buildings where classes are scheduled. %

\noindent
\textbf{Participants:}
We recruited participants on a first-come-first-served basis through advertisements and fliers distributed around the university campuses, limited to our inventory of smartwatches. Out of 105 participants who participated for at least 15 days (on average) for the 30-day study, 77 participants completed all their assigned tasks (and thus only their data was used in our analysis in \cref{res}), while the remaining participants did not complete their tasks due to varying reasons, such as loss of interest, damaged sensing hardware, or other technical difficulties. Among the participants who completed their tasks, 41 were female, and 36 were males. Their age ranged between 18 and 54 years, and all of them were either students or employees at the university. 61 of the 77 participants primarily attended classes or worked on the main campus, while 16 attended the downtown campus for one or more classes or work. We renumerated the participants with \$25 for their participation in our data collection program. Our participant recruitment, data collection, and result dissemination procedures were reviewed and approved by the university's Institutional Review Board (IRB).

\noindent
\textbf{Sensing Hardware and Application:} 
In order to capture BLE packets broadcast by the e-scooters and at the same time collect additional physiological and contextual information related to each encounter, we loaned a smartwatch to each participant for the duration of their participation. The loaned watch came installed with a custom sensing and data collection application written by us, and was paired with the participant's smartphone only for Internet connectivity (in order to upload the sensed data to our data servers). Only our loaned smartwatch hardware and the installed data collection application was used to sense and collect data. This was done to maintain data consistency (across participants), ease of application development (only one mobile OS and hardware were needed), to avoid liability due to damaging participants' device, and for improving accessibility of carrying out some of the manual tasks (described below) during each encounter. We chose the state-of-the-art Mobvoi TicWatch E smartwatch as our data collection because of its built-in GPS and heart rate sensors, modern BLE v4.1 radio, and IP67 rated water resistance. The TicWatch E also features a 1.4 inch round OLED display and runs Wear OS based on Android 8.0.

\begin{figure}[]
\begin{subfigure}[t]{0.32\linewidth}
\centering
\includegraphics[width=\textwidth]{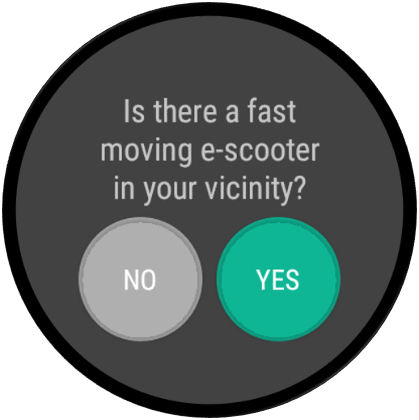} 
\caption{}
\label{}
\end{subfigure}
\hfill
\begin{subfigure}[t]{0.32\linewidth}
\centering
\includegraphics[width=\textwidth]{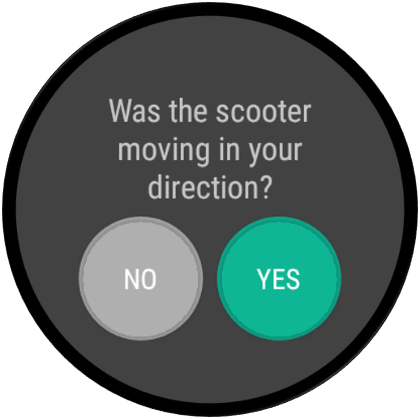} 
\caption{}
\label{}
\end{subfigure}
\hfill
\begin{subfigure}[t]{0.32\linewidth}
\centering
\includegraphics[width=\textwidth]{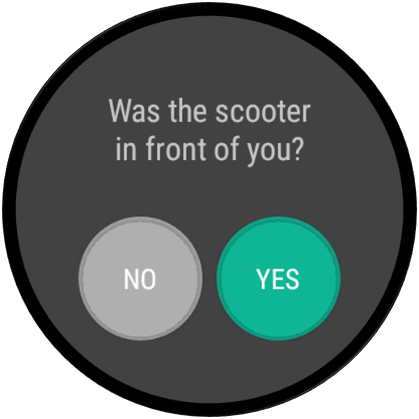} 
\caption{}
\label{}
\end{subfigure}
\caption{Encounter questions.}
\label{fig:encounter-questions}
\end{figure}

\noindent
\textbf{Participant Tasks:} 
Each participant was required to wear the loaned smartwatch, especially when present on any one of the university campuses, for a total of at least 30 days. We initiated the data collection program in April 2019 and terminated it by the end of June 2019 (a total of 3 months). On the first day of participation, participants signed the IRB-approved consent form, completed a demographic survey, checked out the smartwatch with the installed data collection application, received assistance in pairing the loaned smartwatch with their phones and received a brief orientation on the operation of the installed application and their expected tasks. Whenever our data collection application (running in the background) determines\footnote{Accomplished using Android's \texttt{\small{DetectedActivity}} API.} that the participant is a pedestrian and if any e-scooter is detected in their vicinity (i.e., by sensing the BLE packets originating from the e-scooters) at that time, it prompts the participant to answer up to three Yes/No questions (\cref{fig:encounter-questions}) related to the encounter. The goal of these questions is to collect some real-time ground truth related to the detected encounter. If the participant answered \emph{NO} to the first question (\emph{``Is there a fast moving e-scooter in your vicinity?''}), the remaining two questions related to the e-scooter mobility were not asked. If participants failed to answer the questions within a short period (say, within a minute) after the e-scooter detection, the interface displaying the question was no longer available to prevent false data entry. In order to prevent annoyance to participants, and to preserve participant engagement throughout the data collection period, the application asked questions only once every 15 minutes, even if the participants encountered more than one e-scooter during that period. Also, during the first-day orientation, participants were instructed that they can be as engaged in providing real-time feedback as they want, removing any pressure or coercion for providing feedback. On their last day of participation, participants returned the loaned smartwatch (and any other accessory), completed a post-study pedestrian safety survey, and got remunerated. Details of the post-study survey instrument and its outcomes are presented later in \cref{subsec:perceptions}.

\subsection{Data Modalities} \label{data-modalities}
We collected real-time quantitative data related to the encounters between e-scooters and our participants employing the data collection procedure and application described above. \cref{tab:data-modalities} summarizes all the information or data related to these encounters that were either directly sensed or indirectly inferred. Due to moderate weather conditions throughout the study period, with an average temperature of 74.72$^\circ F$ ($\sigma=$ 4.87$^\circ F$) and average precipitation of just 0.137 $in.$ per month ($\sigma=$ 0.03 $in.$) \cite{wunderground_2020}, we deem that our dataset will not be very useful for understanding the impact of weather on e-scooter and pedestrian safety.

{
\renewcommand\arraystretch{1.33}
\begin{table}[h]
\scriptsize
\centering
\caption{List of all information from/about the encounters.}
\begin{tabular}{|l|l|}
\toprule
Quantitative & External \\
\midrule
Location & Pedestrian and Rider Attractors \\
Time & \quad$\bullet$ Location \\
Heart Rate & \quad$\bullet$ Time \\
Bluetooth   & Pedestrian and Rider Generators \quad \\
\quad$\bullet$ Signal Strength & \quad$\bullet$ Location \\
\quad$\bullet$ Service Provider & \quad$\bullet$ Time \\
On-Spot Questions & \\
\quad$\bullet$ Stationary or Moving & \\
\quad$\bullet$ In Front or Behind & \\
\quad$\bullet$ Direction w.r.t. Pedestrian \quad &  \\
\bottomrule
\end{tabular}
\label{tab:data-modalities}
\end{table}
}

\noindent
\textbf{Quantitative Data:}
Our data collection application logged participants' every encounter with e-scooters in their vicinity. Specifically, it recorded the signal strength information from the BLE packets received from the e-scooter(s), time, location (GPS coordinates), heart rate, and participants' responses to the three questions (\cref{fig:encounter-questions}) if available. By conducting a comprehensive heuristic analysis of the BLE advertisement packets before the start of the study, we determined a technique for identifying the service provider corresponding to each received BLE packet. Using this information, our data collection application also recorded the service provider corresponding to each encountered e-scooter.  

\noindent
\textbf{External Data:}
We also collect certain supplementary information that can help us understand and/or support our findings from the quantitative data. Specifically, we gathered location and time information on pedestrian and rider \emph{attractors} and \emph{generators}. We refer to locations where a significant number of pedestrians and riders are headed, such as a class starting at a particular time, as attractors. Similarly, generators are locations where a significant number of pedestrians and riders are generated, such as a bus stop or parking lot. Attractors and generators often play dual roles, for example, when a class ends and another starts just afterward. We collectively refer to such attractors and generators as \emph{points of interest (or POI)}. 

\subsection{Encounters and Data Sources}
\label{encounter-definition}
An \emph{encounter}, as relevant to our analyses, occurs when an e-scooter and a pedestrian meet each other at close proximity. Detecting such encounters from our crowd-sensed data is important, and a prerequisite, before analyzing their spatio-temporal characteristics for safeness. Thus, we first define the notion of an encounter based on available data (BLE and user feedback) as follows: 
\begin{itemize}[leftmargin=*]
\item \textbf{Predicted Encounters ($E_P$)}: Derived from BLE data and tagged by the $algorithm$ in \cref{pred} after the study.
\item \textbf{Observed Encounters ($E_O$)}: Derived from feedback data and tagged by the $participant$ in real-time during the study.
\end{itemize}

While $E_P$ is more deterministic, $E_O$ has information about the direction and location of the e-scooter with respect to the pedestrian and thus can provide safety insights on moving e-scooters based on user feedback (\cref{tab:data-modalities}). We provide details of both encounters in \cref{pred,tagged}, respectively.

\begin{figure}[]
\centering
\includegraphics[width= \linewidth, height=6cm, keepaspectratio]{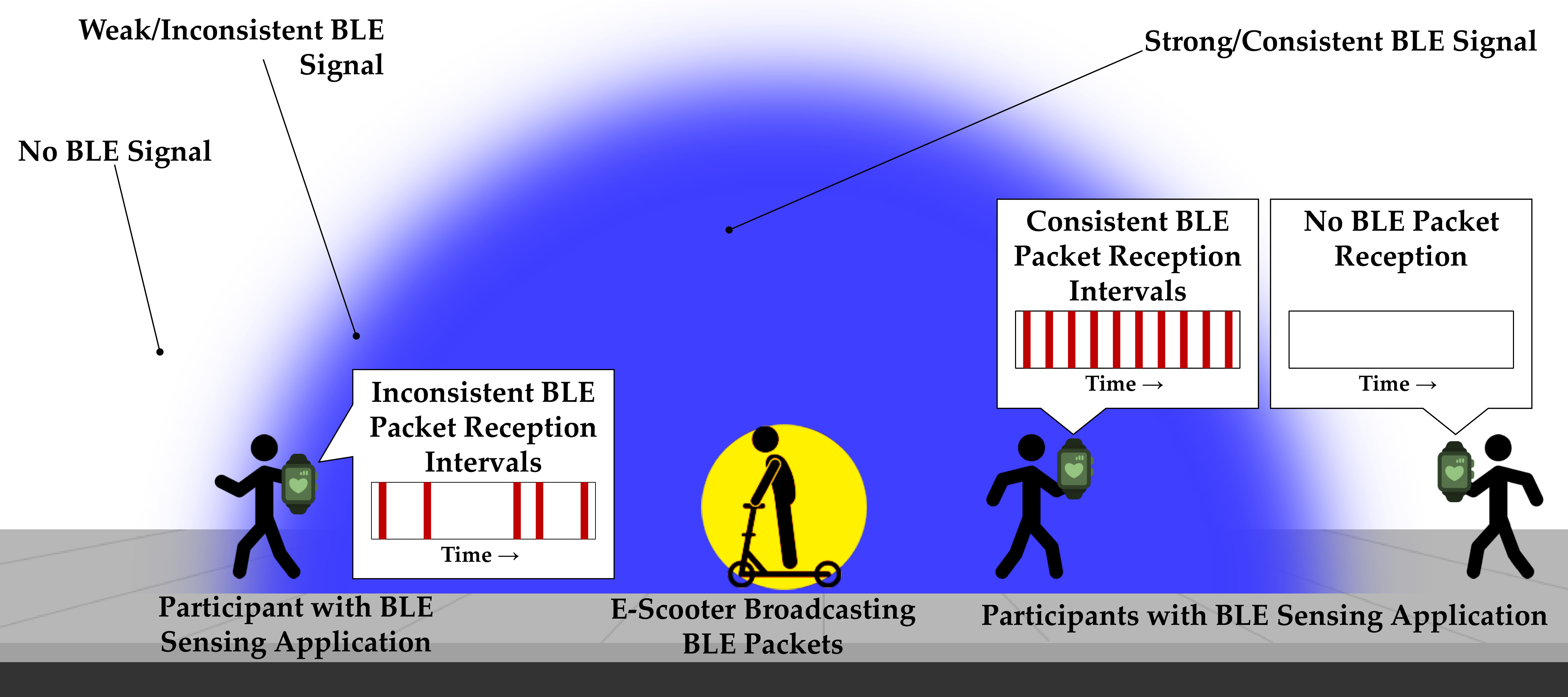}
\caption{BLE signal coverage around an e-scooter and how pedestrians at different distances from the e-scooter observe different reception intervals between BLE advertisements.}
\label{fig:ble-signal-strength}
\end{figure}

{\parindent0pt
\subsubsection{$E_P$}
\label{pred}
Data packets broadcast by BLE radios on-board e-scooters is a reliable means to determine proximity between participants and e-scooters, however not all close-enough encounters may be relevant to our analysis. For example, our participant could have captured one or two BLE packets from inside their home when an e-scooter rode past their house, which should not be considered as a real encounter. Prior to our field study, we empirically determined that as a pedestrian moves away from an e-scooter (i.e., the distance between them increases), reception intervals of the BLE packets transmitted by the e-scooter becomes inconsistent at his/her smartwatch (as shown in \cref{fig:ble-signal-strength}). For instance, we start observing inconsistent BLE reception intervals, starting at a distance of 20-25 $ft$ or more. This observation was consistently observed across most models of the e-scooters and providers targeted by us in this work. We use this observation to classify a sequence of captured BLE packets as an encounter and then outlined an efficient technique to detect such encounters within the BLE packet stream in our dataset.
}

\begin{figure}[b]
\centering
\includegraphics[width= \linewidth]{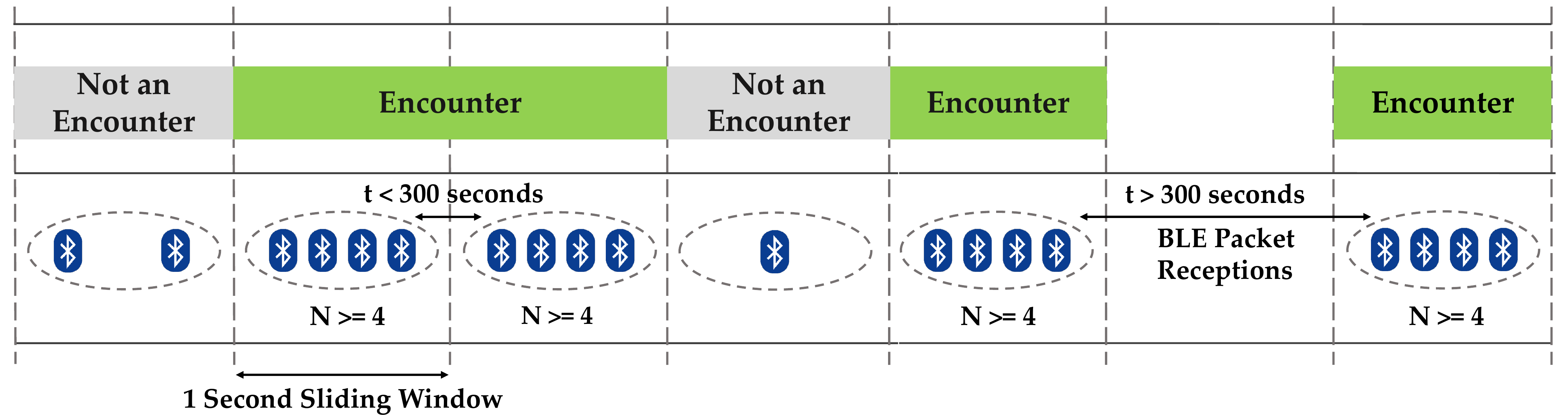}
\caption{Encounter detection algorithm on different BLE reception cases.}
\label{fig:encounter-detection}
\end{figure}

We use a \textit{sliding window} approach to identify the encounters in a stream of fragmented BLE packets captured throughout the day by each participant. A \textit{window size} of 1 $second$ with an 80\% overlap (i.e., each window has an overlap of 80\% with its previous window) is used, and the windows that contain 4 or more BLE packets are marked as \emph{potential} encounter windows. Both the window length and threshold of 4 were empirically determined, based on the approximate minimum encounter duration and approximate maximum BLE advertisement interval, respectively. The potential encounter windows are then further refined as follows: If the time interval of BLE packets between two (or more) potential encounter windows is less than 300 $seconds$, the two windows are combined to form a single encounter. If the time interval is greater than 300 $seconds$, the two windows are considered as two separate encounters. Finally, if more than 4 encounters are detected for a specific e-scooter in one day by a single participant, the later encounters are discarded on the basis that the participant was spending abnormally long durations of time in close proximity of the same e-scooter (e.g., sitting at an outdoor restaurant where a scooter was parked nearby). This filtering step ensures that a single e-scooter or participant does not heavily bias our encounter data and the related analysis. \cref{fig:encounter-detection} summarizes this encounter detection technique with different BLE packet reception examples. Using the above encounter detection technique, we classified e-scooter--pedestrian encounters with 1058 of the 7919 uniquely observed  e-scooters in our dataset (determined using unique identifiers in the captured BLE packets). Overall, we observed a total of 1800 predicted encounters, including repeat encounters with previously encountered e-scooters.

\begin{figure}[]
\includegraphics[width=0.95\linewidth]{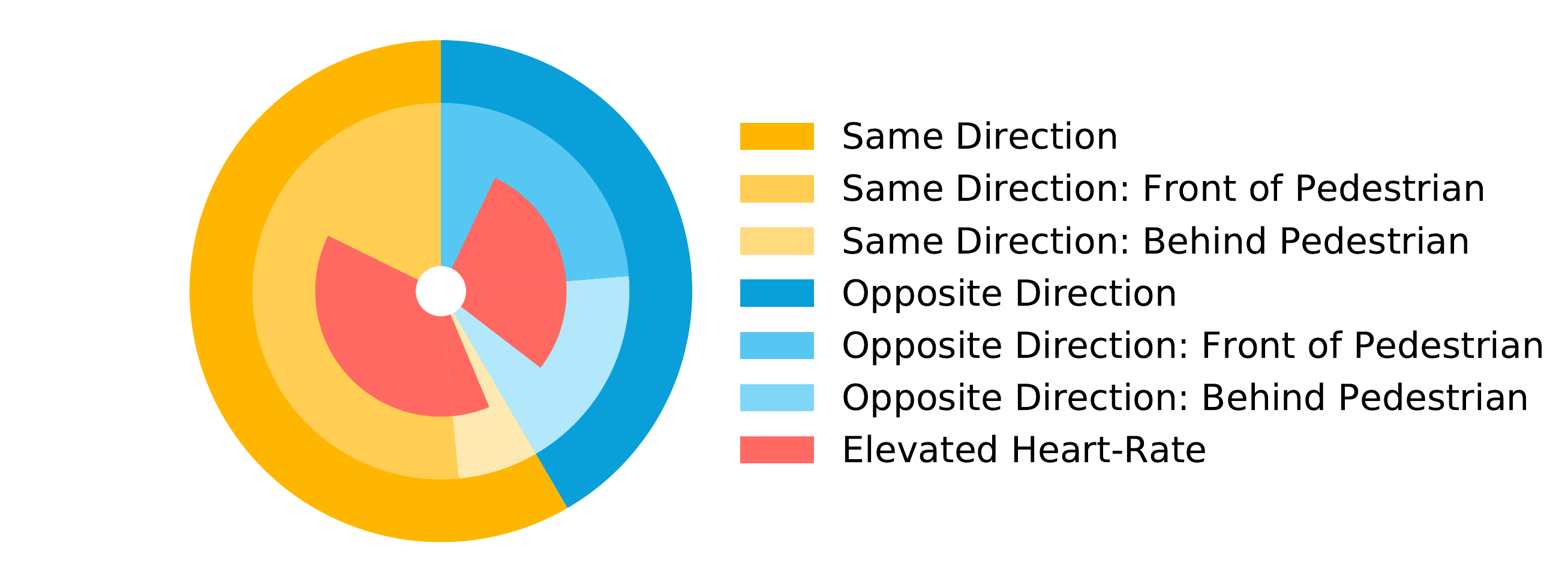}
\caption{Summary of observed ($E_O$) encounters for e-scooter moving direction and pedestrian line-of-sight combinations.}
\label{fig:obs-enc-summary}
\end{figure}

{\parindent0pt
\subsubsection{$E_O$}
\label{tagged}
Observed encounters are voluntarily tagged by the participant in real-time whenever the data collection application detects an e-scooter in proximity. In contrast to passively collecting the BLE data, which does not require active involvement of the participant, reliable feedback data is challenging to collect because it not only requires reliance on the participants to actively provide feedback, but such data is also subjective. This aspect was identified in our dataset, which contains 6482 feedbacks (among 10000+ detections) on e-scooters over the entire study period. In both encounter types ($E_P$ and $E_O$), Blue Duck's e-scooters constituted only 2\% of all detected e-scooters and were not observed in the feedback data. Therefore, we will use only encounters from Bird and Lime brand of e-scooters for our analysis. Moreover, we will only consider encounters that occur between 06:00-23:00 (4993 feedbacks), because the earliest class (on either campus) started at 07:00 and the last class finished at 21:45. Therefore, the time period between 06:00-23:00 represents the most typical use of e-scooters as a last-mile transportation solution. Approximately 20\% of the recorded observations in that period correspond to moving e-scooters, with at least 100 potentially hazardous observations where the e-scooter approached the participants from behind. A breakdown of the observed ($E_O$) encounters to show the different e-scooter moving direction and pedestrian line-of-sight combinations appears in \cref{fig:obs-enc-summary}.
}

An increase in heart rate can occur when a pedestrian is startled by a fast-moving e-scooter, which in many scenarios implies that the pedestrian was faced with inadequate response time. We study this parameter to validate if our participants were startled by the observed e-scooter encounter or not. For the analysis, we use the heart rate data that was collected from each participant whenever a feedback questionnaire was triggered. The \textit{normal or resting} heart rate can vary significantly from participant to participant, which hinders the feasibility of using a global threshold for all participants. Thus, we determine personalized threshold ranges for each participant based on their overall heart rate data, and their most frequently occurring pulse rate(s). We use this threshold to check if an encounter-related (moving) heart rate was within the participant's computed threshold (for most daily activities) or not. In almost 60\% of the moving encounters seen in \cref{fig:obs-enc-summary}, participants (as pedestrians) have an elevated heart rate with e-scooters approaching them \textit{within one foot away at some time instant} from the front and the behind. This finding aligns with our intuition that pedestrians may have little time to respond to rapidly moving e-scooters and can be easily startled by them. Moreover, most e-scooters emit minimal audible sound during their regular operation and combined with their faster speed they could present a significant safety risk to the pedestrians if they cannot observe them and take appropriate reactions in a timely fashion.

%% file: findings.tex
\section{Empirical Findings}
\label{res}
In this section, we comprehensively analyze the data collected during our field study by employing the criteria outlined in \cref{goals}. 

\begin{figure*}[]
\centering
\includegraphics[width= 0.875\linewidth]{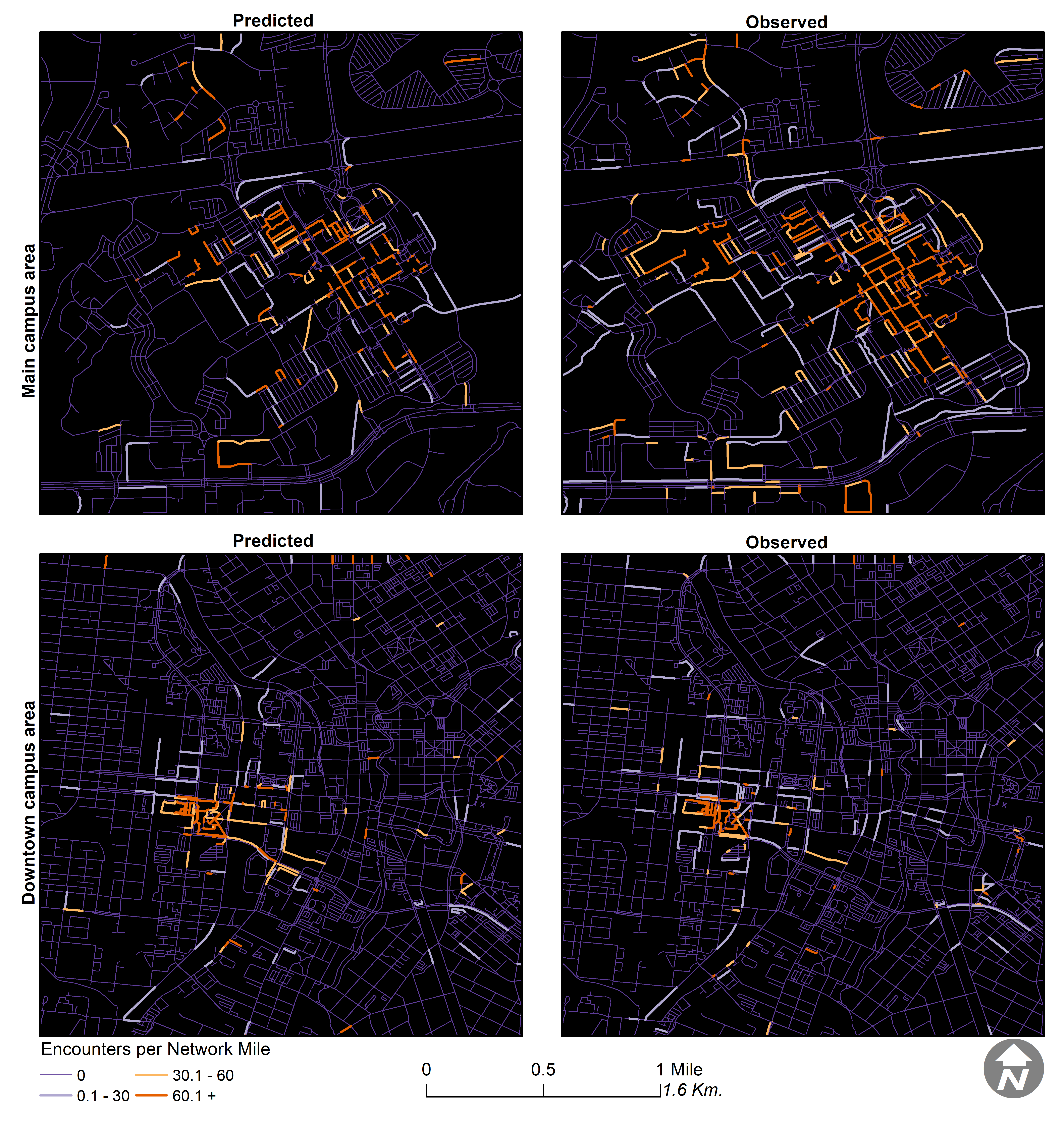}
\caption{Predicted ($E_P$) and observed ($E_O$) encounter density in and around main campus, and downtown campus.}
\label{fig:spatial-char}
\end{figure*}

\subsection{Outcomes of \ref{rq:space}}
\begin{figure*}[]
\centering
\begin{subfigure}{0.32\linewidth}
\centering
\includegraphics[width=\textwidth]{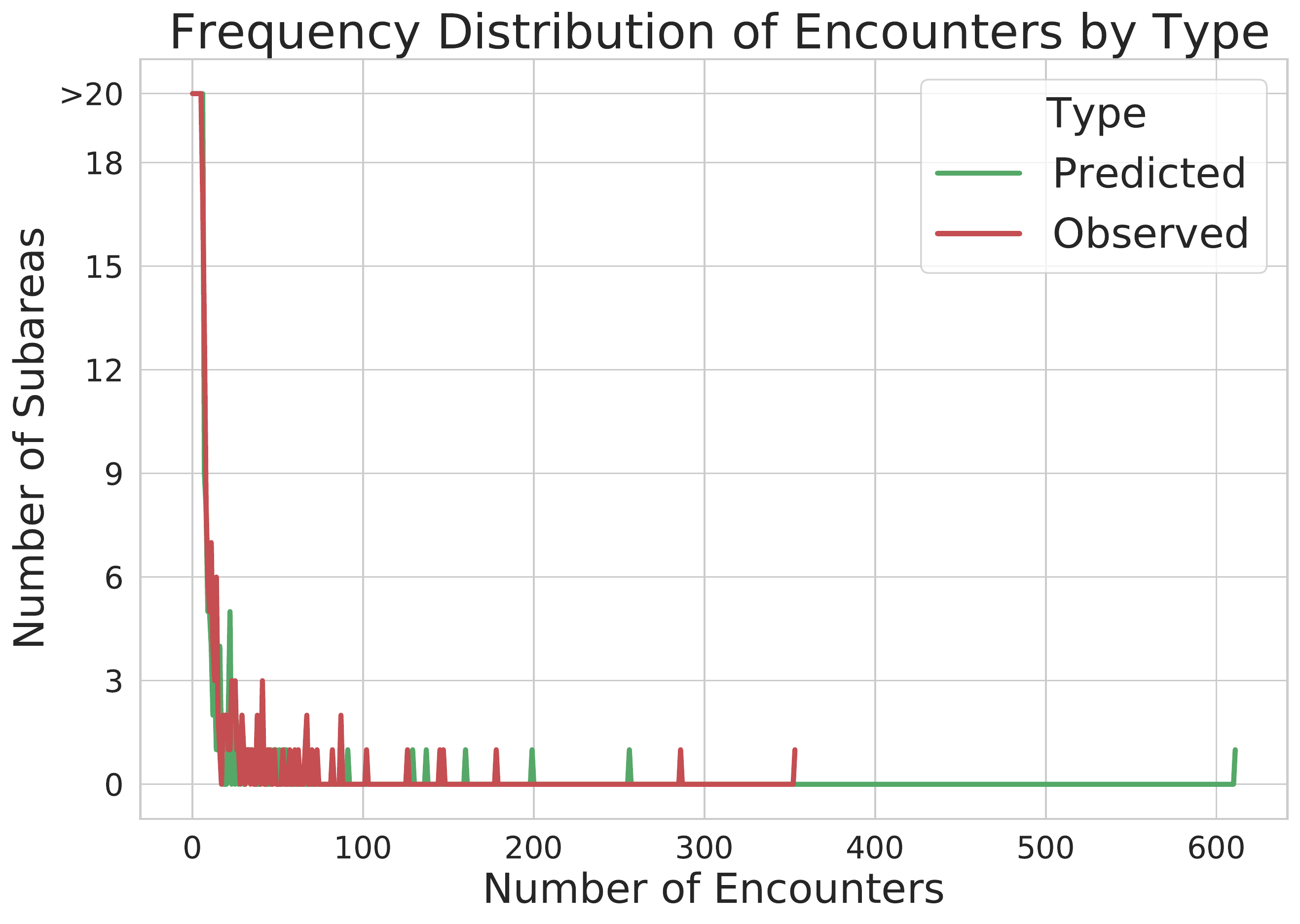}
\caption{Space: Atomic road segments.}
\label{fig:frequency-space}
\end{subfigure}
\begin{subfigure}{0.32\linewidth}
\centering
\includegraphics[width=\textwidth]{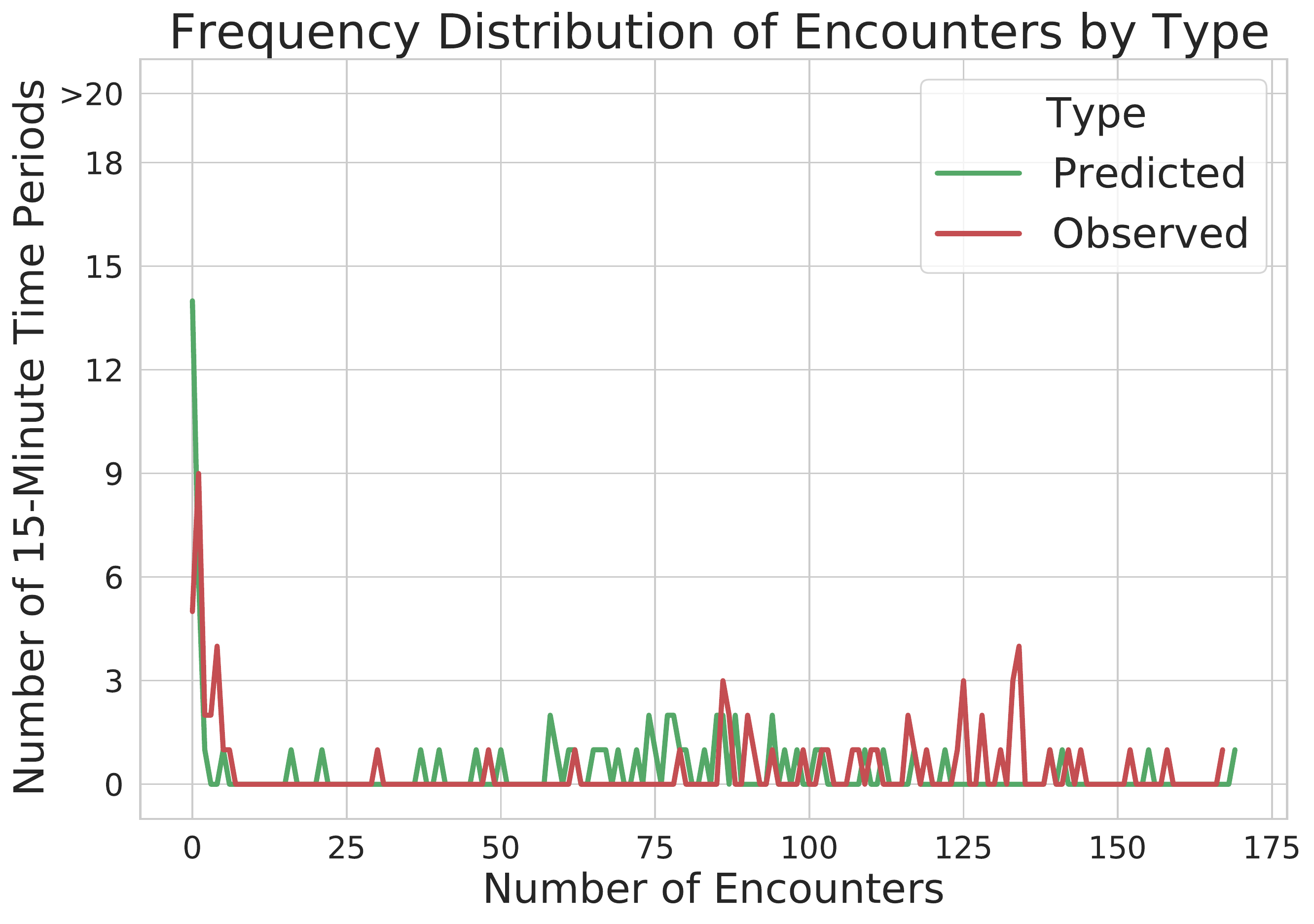}
\caption{Time: 15-minute periods.}
\label{fig:frequency-time}
\end{subfigure}
\begin{subfigure}{0.32\linewidth}
\centering
\includegraphics[width=\textwidth]{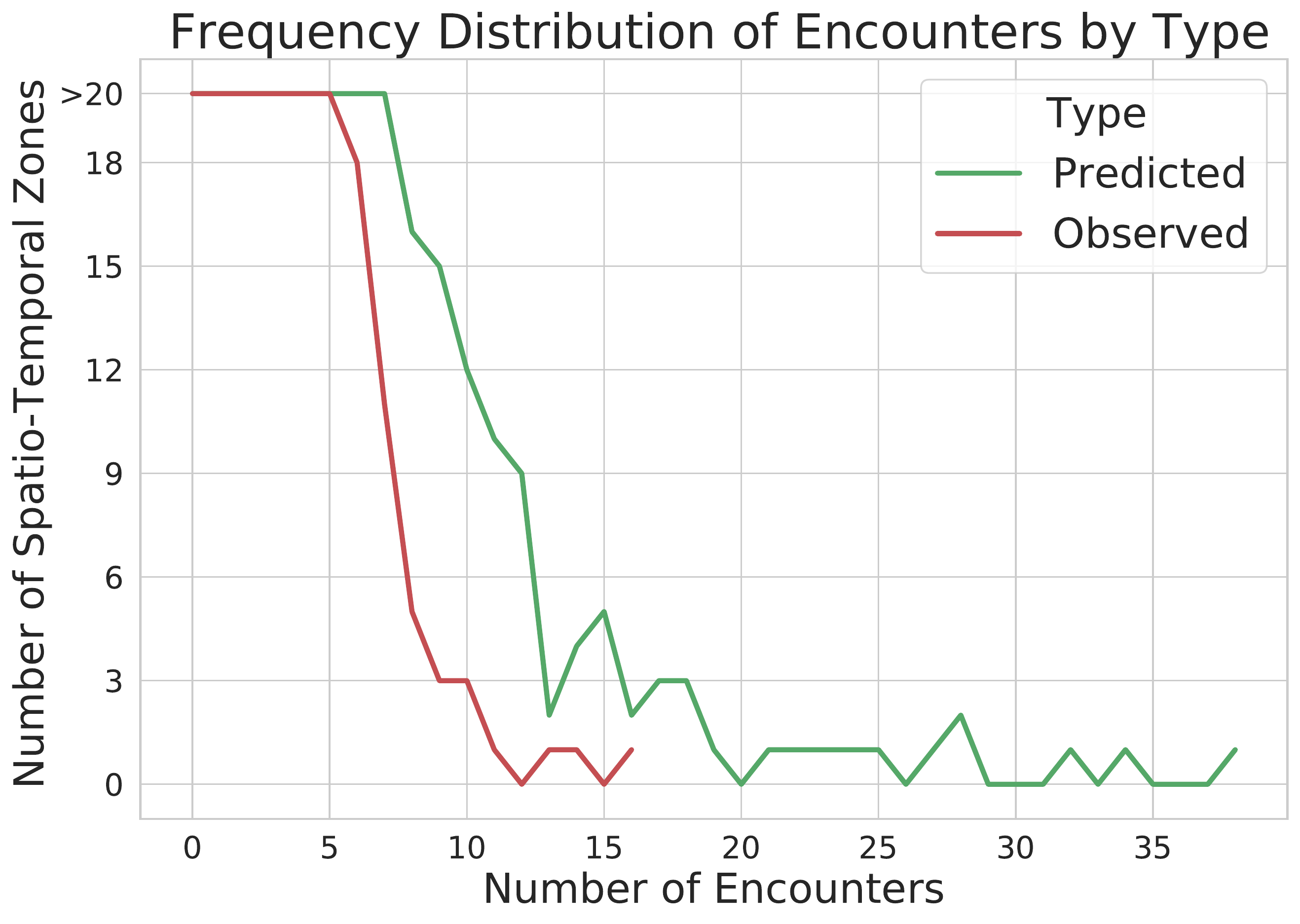}
\caption{Space-Time: All combinations.}
\label{fig:frequency-space-time}
\end{subfigure}
\caption{Frequency distribution of predicted ($E_P$) and observed ($E_O$) encounters between 06:00-23:00 among (a) 21447 atomic segments in main and downtown campuses combined, (b) 68 15-minute periods in a day, and (c) all combinations of 21447 atomic segments and 68 15-minute periods.}
\label{fig:frequency}
\end{figure*}

To analyze how the encounters are spatially distributed throughout our university campuses and their surroundings, we first build a set of \textit{atomic segments} where encounters may occur. Each atomic segment is an edge in the graph of roads and walkways, and one can enter or exit an atomic segment only at its end. Atomic road segments may connect with other atomic segments (such as at an intersection), or end at a POI. An encounter map in \cref{fig:spatial-char} shows the number of predicted  ($E_P$) and observed  ($E_O$) encounters that occurred in each of the campus areas, during the entire study period. The highest encounter counts in the main campus atomic segments are $E_P=611$ and $E_O=35$, whereas in the downtown campus are $E_P=256$ and $E_O=55$. Out of the 21447 atomic segments (combined for both the main and downtown campuses) from \cref{fig:frequency-space}, at least twenty atomic segments in both campuses have a relatively high number of encounters: $E_P>25$ and $E_O>5$, with more than 95\% of atomic segments having five or fewer $E_P$ and $E_O$. These results highlight the extremely disproportionate number of encounters on both campuses, implying that pedestrians in certain parts of the campuses (and their surroundings) are significantly more likely to encounter e-scooters than the rest of the campuses.

From \cref{fig:spatial-char}, we notice a high saturation of e-scooter encounters around student residential areas on both on- and off-campus locations. This saturation occurs as students disperse from the residential areas to different academic buildings of the university where classes are being held. As part of their commute, they may walk for a short distance from these residential areas until they reach a shuttle stop or a parked e-scooter or their final destination. Such daily commutes result in heavy pedestrian traffic near the residential areas, which are often targeted by service providers for deploying their fleet. These factors explain the presence of high encounters ($E_P$ and $E_O$) counts in these areas. We also notice similar saturation at other POIs near shuttle stops and outside staircases at the end of a long walking path. Riders, commuting to buildings inside the university, park the e-scooters near the stairs or outside doorways, as carrying their e-scooters through to the staircase is inconvenient. This scenario can explain the high number of e-scooter encounters ($E_P$ and $E_O$) near these places.

We next focus on the spatial closeness of the predicted encounters because closer encounters have a higher likelihood of resulting in a pedestrian-related collision or disruption. Due to its attenuation over distance, BLE signal strength is a good indicator of the spatial closeness of encounters that occurred between our participants and e-scooters operating in our test deployment area. Because of the different BLE transmission power used by different service providers, we conduct this analysis separately for Bird and Lime. The signal strength of BLE packets captured (on the TicWatch E smartwatch) from Bird brand e-scooters within one feet away from their computer module (usually mounted on the stem of the e-scooter) is approximately -60.5 $dB$, whereas for packets captured from Lime brand e-scooters in the same setting has an approximate signal strength of -46.25 $dB$. Using this baseline observation, we found that 0.43\% of encounters ($E_P$) were less than one foot away from the participant.

\begin{figure}[]
\centering
\includegraphics[width= \linewidth]{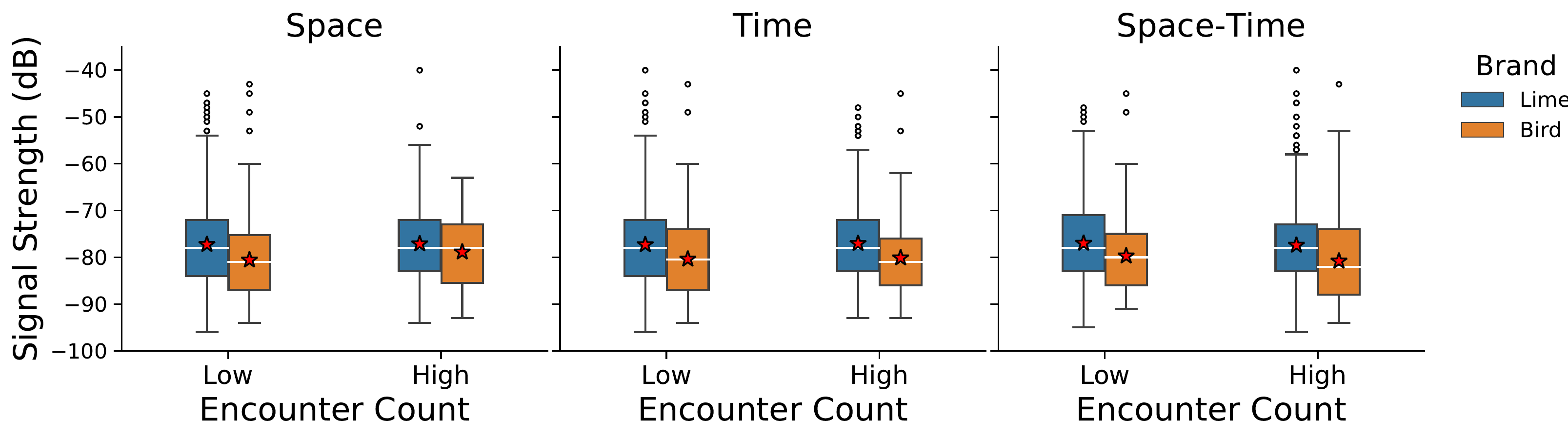}
\caption{Maximum BLE signal strength during each predicted encounter $E_P$. Comparison between atomic segments with low encounter counts (1-84), 15-minute periods and their spatio-temporal combinations vs. atomic segments with high encounter counts (85-169), 15-minute periods and their spatio-temporal combinations, respectively. Star sign denotes the mean BLE signal strength.}
\label{fig:ble-char}
\end{figure}

As seen in \cref{fig:ble-char}, we discovered that predicted encounters in atomic segments with high encounter counts are on average closer (as the average BLE signal strength is relatively stronger) than predicted encounters in atomic segments with low encounter counts (as the average BLE signal strength is relatively weaker). This analysis tells us that encounters in high-encounter atomic segments are at a relatively closer range (distance between the participants and e-scooters) than encounters in low-encounter atomic segments, which indirectly suggests that collisions are more likely to occur in high-encounter atomic segments than in low-encounter atomic segments. %

{
\renewcommand\arraystretch{1.33}
\begin{table}
\scriptsize
\centering
\begin{threeparttable}
\caption{Space: Encounters by functional classification.}
\label{tab:space}
\setlength{\tabcolsep}{5.5pt} %
\renewcommand{\arraystretch}{1.3} %
\centering
\begin{tabular}{|c|c|c|c|c|c|c|} \cline{2-7}
\multicolumn{1}{c|}{} & \multicolumn{2}{c|}{\textbf{\textit{TES\textsuperscript{a}}}} & \multicolumn{2}{c|}{\textbf{\textit{MEM\textsuperscript{b}}}} & \multicolumn{2}{c|}{\textbf{\textit{PEM\textsuperscript{c}}}} \\ \hline
\textbf{Functional Class\textsuperscript{d}}  & $E_P$  & $E_O$  & $E_P$  & $E_O$  & $E_P$  & $E_O$  \\ \hline
Arterial Streets & 998 & 709 & 146.1 & 60.7 & 6.9 & 2.3 \\ \hline
Collector Streets & 269 & 336 & 68.4 & 55.2 & 3.2 & 2.1 \\ \hline
Local Streets & 1285 & 2255 & 176.0 & 171.8 & 8.3 & 6.6 \\ \hline
Shared-use Paths & 102 & 119 & 306.0 & 432.6 & 14.5 & 16.6 \\ \hline
Sidewalks & 994 & 1163 & 617.8 & 470.7 & 29.2 & 18.1 \\ \hline
Other/Unclassified & 154 & 411 & 799.1 & 1410.0 & 37.8 & 54.2 \\ \hline
\textbf{Total } & 3802 & 4993 & 352.2 & 433.5 & 100.0 & 100.0 \\ \hline
\end{tabular}
\begin{tablenotes}
\item[a] Total Encounters per Segment ($TES$) is the sum of all detected proximal pedestrian-scooter encounters in a network segment. 
\item[b] Mean Encounters per Mile ($MEM$) is the average number of encounters per segment divided by the length of the segment in miles.
\item[c] Percent Encounters per Mile ($PEM$) refers to the percentage of $TES$ w.r.t sum total of all encounters over all segments.
\item[d] Arterial streets include OpenStreetMap (OSM) API tags "primary" and "secondary". Collector streets include OSM tags "tertiary". Local streets include OSM tags "residential" and "service". Shared-use paths include OSM tags "path" and "cycleway". Sidewalks include OSM tags "footway" and "pedestrian". Other/unclassified uses all other OSM tags.
\end{tablenotes}
\end{threeparttable}
\end{table}
}

\begin{figure*}[]
\centering
\includegraphics[width= 0.67\linewidth]{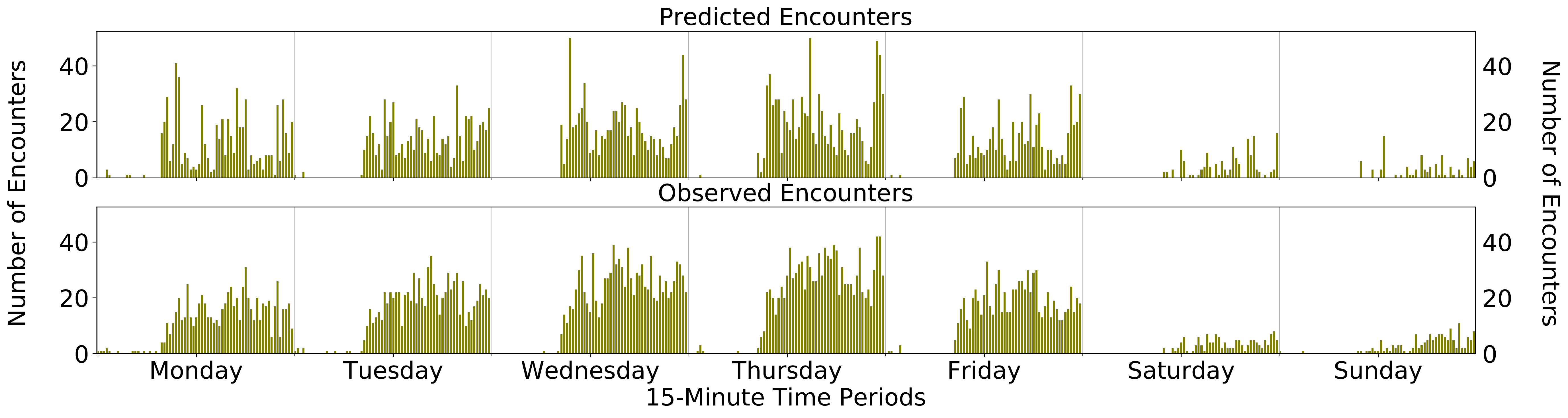}
\caption{Number of predicted ($E_P$) and observed ($E_O$) encounters in each of the 476 15-minute periods (sorted chronologically), between 06:00-23:00 for seven days of a week.}
\label{fig:temporal-char}
\end{figure*}

\begin{figure*}[]
\centering
\includegraphics[width= 0.67\linewidth]{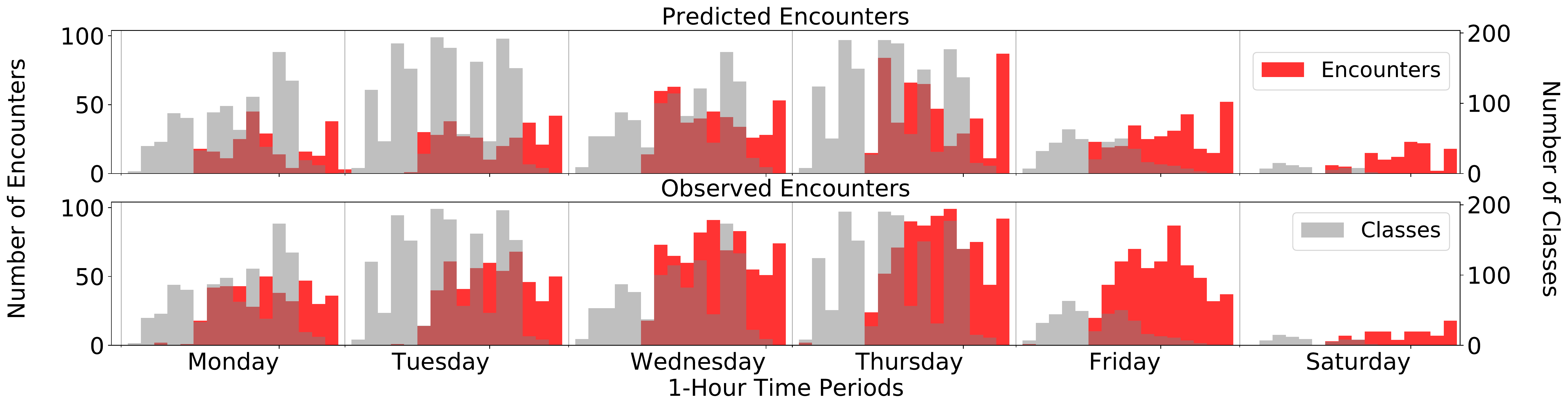}
\caption{Number of predicted ($E_P$) and observed ($E_O$) encounters in each of the 102 1-hour periods (sorted chronologically), between 06:00-23:00 for six days of a week, plotted along with the number of classes scheduled in the corresponding time periods. No regular classes were scheduled on Sundays.}
\label{fig:temporal-diffusion}
\end{figure*}

We also observed that the vast majority of proximate encounters between e-scooter riders and pedestrians happened on narrow pedestrian paths such as sidewalks (\cref{tab:space}). However, there are very few bike lanes and shared-use paths (typically at least 10 $feet$ wide) in the study areas. This deficit creates conflicts and safety challenges for both pedestrians who prefer to walk to nearby buildings, and to riders who may be passing along to reach adjacent parking lots or other destinations. %
Providing infrastructure with separated routes for e-scooters, such as shared-use paths and bike lanes, may help protect pedestrians from conflicts in constrained space on sidewalks. Alternatively, educating all road users on usage guidelines (such as right-of-way and safety rules) via signboards and posters, especially on roadways with high pedestrian and e-scooter density, can prevent future mishaps related to e-scooters. Additionally, \cref{fig:spatial-char} shows several high-encounter roadways in relatively isolated locations. Planners and engineers can review these spots for targeted projects to reduce conflicts.

\noindent
\textbf{Generalized Implications.} These findings suggest opportunities to improve safety both in and outside a campus setting. Service provider data can easily identify e-scooter density and rider routes around a specific area at a given time. In contrast, our data allowed identifying pedestrian-scooter encounters and their density in specific areas, for instance, parking lots, recreation centers, etc. With this knowledge, planners can identify these hotspot areas and remediate areas that lack adequate critical infrastructure through rules of co-existence (redirecting flow, rearranging) or additional structures (lanes, docking stations). For instance, a safe walking space could be reclaimed in high-encounter areas by adding a bike lane alongside paths for micromobility and bicycle riding, or along nearby parallel routes \cite{nacto,mckenzie2020urban}. Furthermore, the above results can help optimize transit in a way that will reduce the average distance traveled by last-mile commuters, and thus reduce the number of encounter.

\subsection{Outcomes of \ref{rq:time}}
To analyze how the encounters are temporally distributed throughout the week, we partitioned the week into 476 15-minute periods starting at 06:00 and ending at 23:00 each day. Although there were significantly fewer classes and events on campus during the weekends, we included them in our analysis for completeness. \cref{fig:temporal-char} shows the number of encounters that occurred in each of the 476 time periods across both the campuses, during the entire study period. During two time periods, Wednesdays 12:45-13:00 and Thursdays 22:30-22:45, we observe the highest number of predicted encounters ($E_P>50$). At the same time, for twenty time periods on Wednesdays and Thursdays, we also see a relatively higher number of observed encounters ($E_O>35$). Also, we see several spikes and surges throughout Monday to Friday, and both $E_P$ and $E_O$ on campus were significantly lower on Saturdays and Sundays. To understand the encounter frequency throughout the length of a day, we added the encounter counts observed during the 68 15-minute periods each day (between 06:00-23:00). The results shown in \cref{fig:frequency-time} demonstrates that pedestrians are significantly more likely to encounter e-scooters at certain times of the day, such as between 12:45-13:00 and between 14:45-15:00. During these time slots, our participants had a total of $E_P=169$ and $E_O=28$.

\begin{figure*}[h]
\centering
\begin{subfigure}{0.44\linewidth}
\centering
\includegraphics[width=\textwidth]{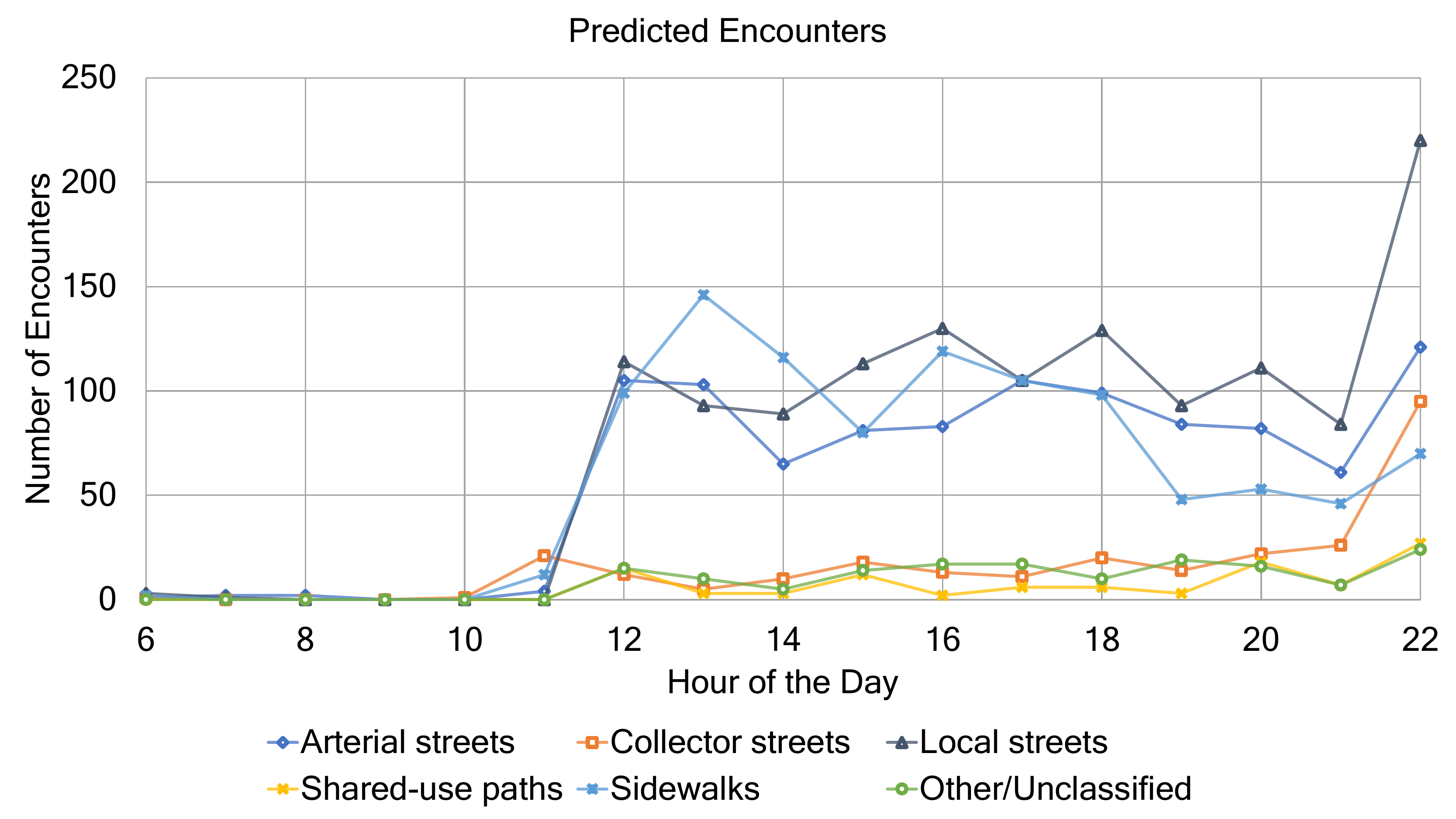}
\label{fig:space-time-observed}
\end{subfigure}
\begin{subfigure}{0.44\linewidth}
\centering
\includegraphics[width=\textwidth]{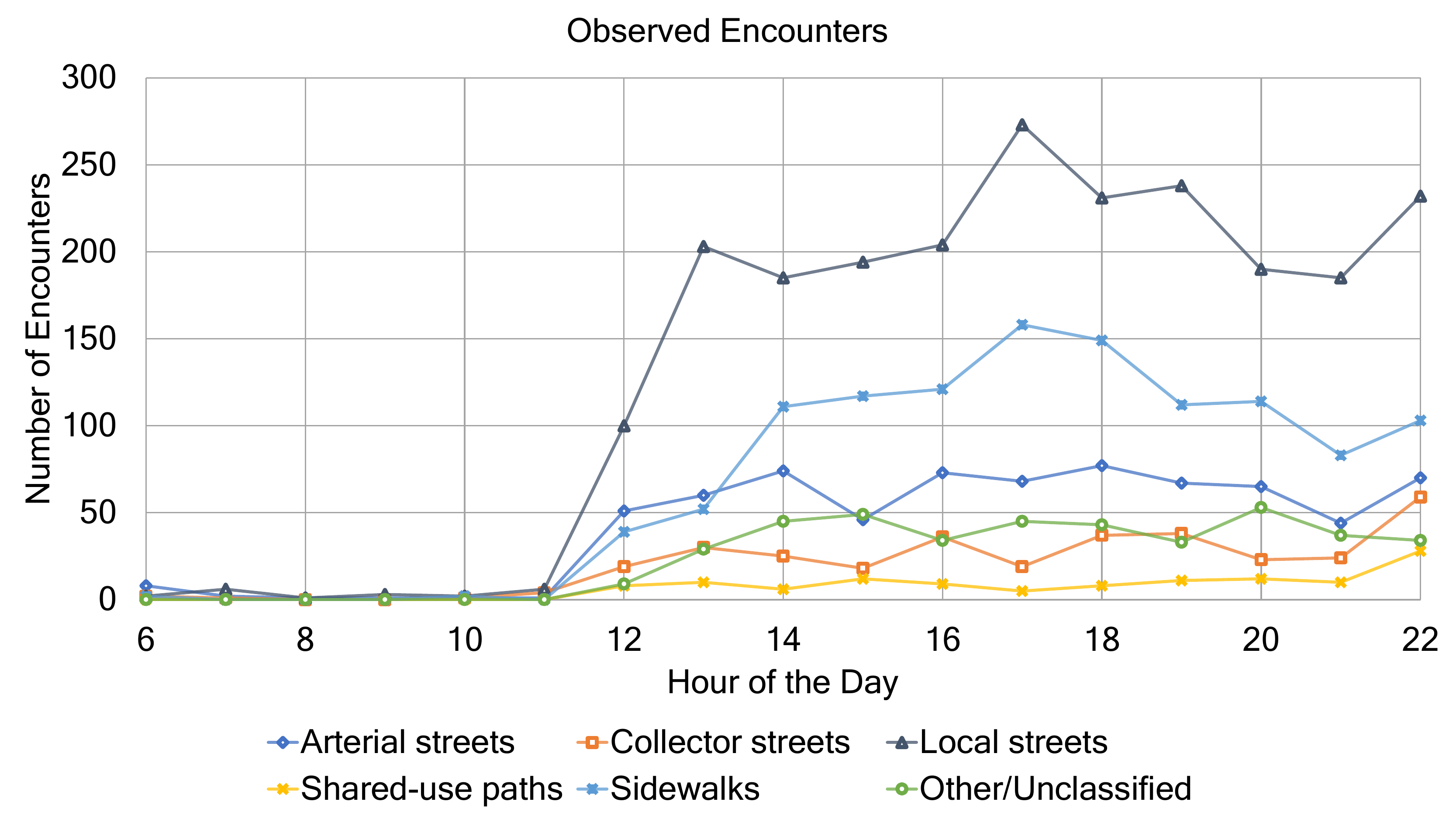}
\label{fig:space-time-predicted}
\end{subfigure}
\caption{Number of predicted ($E_P$) and observed ($E_O$) encounters in each 1-hour time period between 06:00-23:00, plotted for each functional classification of road network segments. The $x$-axis unit represents the next 1-hour time period.}
\label{fig:space-time}
\end{figure*}

Similar to the spatial analysis, we investigate the spatial closeness of encounters that occurred during periods with low encounter counts, and with encounters that occurred during periods with high encounter counts. We split the encounters equally among two interval groups based on the maximum encounter count for each encounter type. As shown in \cref{fig:ble-char}, encounters that occurred during periods with high encounter counts are generally closer, for both the Bird and Lime brand e-scooters, as the observed average BLE signal strength is relatively stronger in these encounters. This finding is in contrast to encounters that occurred during periods with low encounter counts, as the observed average BLE signal strength for encounters is relatively weaker in this case. This distinction suggests that collisions are more likely to occur during time periods with high encounter counts than during time periods with low encounter counts. Similar patterns were observed in both predicted ($E_P$) and observed ($E_O$) encounters.

As students and some employees plan their arrival and departure to/from campus depending on class timings, it is intuitive that our encounter observations have some relation to the schedule of classes. We plot the hourly encounters recorded from April to early-May (when the classes ended at the end of spring semester) alongside the number of classes scheduled per week in \cref{fig:temporal-diffusion}. We observed that the highest number of classes occur on Tuesdays, Thursdays, and Wednesdays in the week in \cref{fig:temporal-diffusion}, and the average encounters these days are also higher than the rest of the week showing the occurrence of encounters follows closely with class schedules. Also, there are more predicted encounters ($E_P$) at night than during the day, more likely due to late-night study and exam preparations by students. While we see significant overlap in the afternoons, there are comparatively fewer encounters (predicted and observed) around the early morning periods. This overlap could be due to a combination of multiple factors. First, the personnel who recharge the e-scooters (in return for a payment from the service provider) usually do so during the night, as specific e-scooter models can take up to 8 hours to fully recharge. These personnel generally collect drained e-scooters around late evening or night and are also responsible for distributing the recharged e-scooters around the city. We observed that the recharged e-scooters are usually distributed around the late morning periods, which aligns with our observation of the negligible number of encounters around the early morning periods. Second, late spring-early summer mornings in our target field of study usually have a pleasant climate, which may prompt last-mile commuters to walk to their final destination instead of using micromobility vehicles.

\noindent
\textbf{Generalized Implications.} Scooters can be introduced or removed around the time periods with high encounters to provide reliable transportation options (to reach destinations in a timely fashion without hindering other road users). Regulations can be set accordingly to improve the road user experience and provide a safer environment (for pedestrians) through better management of chaotic times. For example, in a university setting, the shuttle buses can be made more frequent during the observed high encounter times, which could encourage students to use shuttles instead of e-scooters. Similarly, in a crowded city setting, the timing of frequent e-scooter encounters could be used in combination with other travel modes to compliment last-mile connections and reduce conflicts.

\subsection{Outcomes of \ref{rq:space-time}}
To analyze how the observed encounters ($E_O$) are spatio-temporally distributed, we study all combinations of the 21,447 atomic segments in both campuses and 68 15-minute periods in one day (between 06:00-23:00), for a total of 1,458,396 spatio-temporal zones in each campus. More than 90\% of the spatio-temporal zones in both the main and the downtown campus did not have any predicted encounters ($E_P$) or observed encounters ($E_O$), as seen in \cref{fig:frequency-space-time}. This asymmetry indicates that pedestrians are significantly more likely to encounter e-scooters in certain parts of the campuses (and their surroundings) than the rest of the campus areas, and only at specific times. For instance, there were lesser or no predicted encounters ($E_P$) on the Main campus from 06:00-11:00 on Tuesdays, compared to the latter half of the day.

We also identified that the residential areas outside the campuses had fewer or no encounters in the early morning, more likely due to e-scooter recharges schedules, lack of classes, and availability of bus shuttles. We noticed high predicted encounters ($E_P$) inside the campus, mostly between 12:00-14:00. High encounter counts, both predicted ($E_P$) and observed ($E_O$), could be explained by the following factors. Firstly, people usually leave university for lunch around this time. Also, students who only have morning classes for the day start leaving the campus, and on the other hand, students who only have afternoon classes start coming on to campus around this time. Since there are more classes (150+) from mid-day to early-evening (12:00-16:00) on most weekdays except Fridays, depicted in \cref{fig:temporal-diffusion}, this could also show how e-scooter usage also increases around that time, with the highest number of encounters occurring between 14:45-15:00. 

Distribution of encounters by functional classification of their locations in \cref{fig:space-time} shows observed encounters vary more on an hourly basis than predicted encounters. Local street encounters peak mid-day and at 17:00, suggesting an increased interaction with pedestrians during lunch breaks and commuting. Many local streets in the study area do not include sidewalks, which may exacerbate these conflicts. Afternoon and evening peaks in observed encounters using sidewalks suggest their important role in class changes and last-mile connections, yet with limited space to separate pedestrians and e-scooter riders. The low overall conflicts on shared-use paths and other/unclassified network links are likely due to both the improved space for separating modes, and the lower availability of these network links serving destinations. The late evening peak in encounters, particularly on local streets, could be related to high usage of e-scooters for recreational and social trips.

Similar to the individual spatial and temporal analyses outlined earlier, we discovered from \cref{fig:ble-char} that predicted encounters in atomic segments with high encounter counts are on average closer in range (as the observed average signal strength of the BLE packets in the encounters is relatively stronger) than predicted encounters in atomic segments with low encounter counts (as the observed average signal strength of the BLE packets in the encounters is relatively weaker) for Lime and Bird brand e-scooters. This suggests that e-scooter related pedestrian collisions are more likely to occur in spatio-temporal zones with high encounter counts than in the ones with low encounter counts.

\noindent
\textbf{Generalized Implications.} 
The spatio-temporal analysis provides insights on multiple location-time combinations, and can support reduction of conflicts, in addition to multi-modal coordination. Planners and engineers can use encounter data to identify locations for infrastructure improvements that are sensitive to local transportation demands throughout the day. Traffic signal timing and intersection designs may be adjusted to reduce conflicts with pedestrians, including introduction of bicycle boxes (painted spaces in front of vehicle traffic) and bicycle signal heads, in states where e-scooters are regulated similar to bicycles. Transit planners can involve e-scooter providers in changes to schedules and stop locations, to improve last-mile connectivity and predictability for riders. Space-time coordination may be more critical for special events and in separated land uses such as university settings, as compared with mixed-use settings with activities spread throughout the night and day.

%% file: discussion.tex
\section{Discussion}
\label{discussion}

\subsection{Broader Impact and Limitations} 
This approach analyzes both automated and manual pedestrian, and e-scooter interaction data shows new opportunities for supporting safety with emerging travel modes. A review of previous studies showed a need for studies that crowdsource safety information that is missed by police collision records. %
Our approach may be adaptable to support practical pedestrian safety, such as through the development of a real-time collision warning system. However, our study is not without limitations. One main limitation of our field study was that its scope was restricted to the two suburban and urban campuses and surrounding neighborhoods of one university. Subsequently, some of the results and insights gained from the study may be more directly applicable to our university's infrastructure and regulations, locally-available micromobility vehicles, riders, and pedestrians. Our data collection also did not capture encounters with any privately owned e-scooters as most of them do not emit periodic BLE packets. Our results may also suffer from sampling bias due to the low number of participants and the specific recruit channels used. That being said, our first-of-a-kind study's methodology and analyses (including the employed statistics and benchmarks) can serve as a blueprint on how crowd-sensed micromobility data can be used to enable similar safety-related studies in other urban communities. Due to the privacy-sensitive nature of the location data collected in this study, our dataset will be made available only to researchers upon their request.

An essential aspect of our study is that it only relies on participating pedestrians to passively crowd-sense the e-scooter encounters and other sensor data on-board their mobile devices. Although such an approach has its advantages (e.g., relatively low study deployment cost), if this data is supplemented with sensor data collected from the e-scooters themselves, for example, video feed from a camera mounted on the vehicles or packets received by the vehicles' BLE receivers, it could result in an even better analysis. However, employing commercially-operated e-scooters to collect data is not easy due to restrictions put in place by the service providers owning these vehicles. Researchers could deploy their own micromobility vehicles testbed for this purpose, which could enable a much easier data collection process, but deployment and maintenance of such a testbed would be much more expensive.

\subsection{Participants' Perception about Safety}
\label{subsec:perceptions}
Our study ended with participants completing a post-study survey (outlined in Appendix A). In contrast to the quantitative encounter data which helped us gain useful insight on how riders' mobility patterns and infrastructure-related constraints within shared spatio-temporal zones or spaces could impact pedestrians' safety, the post-study survey response data from participants will shed light on their subjective perception of this issue. This survey comprised of two parts: a set of questions to measure how well our data collection application (specifically, the notifications and the manual feedback interface) performed, and another set of questions to capture participants' interests and preferences vis-\`{a}-vis pedestrian safety and mobile device based safety applications. One highlight of responses to the first set of questions is that despite having a minimum 15-minute interval between sending e-scooter detection notifications, one-third of all participants found these notifications annoying. This observation is not surprising as there are many HCI studies \cite{sahami2014large,mehrotra2016my} that show that notifications have a very high chance of causing annoyance if they are not well-designed, and could eventually disengage users. Fortunately, as our application was passively collecting encounter data irrespective of whether participants responded to or ignored our notifications, it did not impact our data collection process significantly. 
Another highlight is in the responses received to our second set of questions, where 58\% of the participants expressed interest in a mobile application that would alert them about potential encounters with electric- or e-scooters. Although this shows that there is significant interest among users to protect their safety from upcoming micromobility transportation vehicles, it also shows that a significant number of users (42\%) are either not interested in such an application or are indifferent to the problem of pedestrian safety from such vehicles. Given the number of participants who were annoyed at the frequency of notifications (from our application), our hypothesis is that the 42\% of the participants who did not express interest in a micromobility vehicle alerting application responded that way because of their displeasure with the high number of notifications in our data collection application. Although our data collection application was not a pedestrian safety application (because it notified users of all encounters and not only the hazardous ones), the above results highlight an important property that any safety application should possess -- a good balance between useful functionality and user engagement through carefully designed notifications. 

%% file: conclusion.tex
\section{Conclusion}
We conducted a field study on crowd-sensing encounter data between e-scooters and pedestrian participants on two distinct urban university campuses over a three-month period. We analyzed specific spatio-temporal metrics and used them as benchmarks to understand the impact on pedestrian safety from e-scooter services. Our analysis uncovered encounter statistics, mobility trends and hotspots which were then used to identify potentially unsafe spatio-temporal zones for pedestrians. We also speculate planning and infrastructure improvements that may help reduce the number of unsafe spatio-temporal zones. Our work provides a preliminary blueprint on how crowd-sensed micromobility data can enable similar safety-related studies in other urban communities.

%% file: appendix.tex
\section*{Appendix A -- Post-Study Survey}

\subsection*{Study Application Usage}

\begin{enumerate}[leftmargin=*]

\item How often did you receive the feedback notifications from the study application?{\renewcommand\arraystretch{1.33}\scriptsize
\begin{table}[H]
  \centering
    \begin{tabular}{|c|c|c|c|c|}
     \toprule
     Rarely \quad 1 & 2 & 3 & 4 & 5 \quad Very often \\
     \bottomrule
    \end{tabular}%
\end{table}}%

\item Did you at any point find the feedback alert notifications to be annoying?{\renewcommand\arraystretch{1.33}\scriptsize
\begin{table}[H]
  \centering
    \begin{tabular}{|c|c|}
     \toprule
     Yes & No \\
     \bottomrule
    \end{tabular}%
\end{table}}%

\item If yes, did you turn the notifications off?{\renewcommand\arraystretch{1.33}\scriptsize
\begin{table}[H]
  \centering
    \begin{tabular}{|c|c|}
     \toprule
     Yes & No \\
     \bottomrule
    \end{tabular}%
\end{table}}%

\item How effective was the notification mechanism?{\renewcommand\arraystretch{1.33}\scriptsize
\begin{table}[H]
  \centering
    \begin{tabular}{|c|c|c|c|c|}
     \toprule
     Not effective at all \quad 1 & 2 & 3 & 4 & 5 \quad Very effective \\
     \bottomrule
    \end{tabular}%
\end{table}}%
\end{enumerate}

\subsection* {General Pedestrian Safety}

\begin{enumerate}[leftmargin=*]
\setcounter{enumi}{5}

\item Have you ever used any wearable technology that provides pedestrian safety?{\renewcommand\arraystretch{1.33}\scriptsize
\begin{table}[H]
  \centering
    \begin{tabular}{|c|c|}
     \toprule
     Yes & No \\
     \bottomrule
    \end{tabular}%
\end{table}}%

\item If yes, please specify some.{\renewcommand\arraystretch{1.33}\scriptsize
\begin{table}[H]
  \centering
    \begin{tabular}{|c|}
     \toprule
     \quad \quad \quad \quad \quad \quad \quad \quad \quad \quad \quad \quad \quad \\
     \bottomrule
    \end{tabular}%
\end{table}}%

\item Would you be interested in a smartwatch application that alerts you about electric scooters in the vicinity?{\renewcommand\arraystretch{1.33}\scriptsize
\begin{table}[H]
  \centering
    \begin{tabular}{|c|c|}
     \toprule
     Yes & No \\
     \bottomrule
    \end{tabular}%
\end{table}}%

\item If yes, what type of alert would you suggest for this scenario? Select all that apply.{\renewcommand\arraystretch{1.33}\scriptsize
\begin{table}[H]
  \centering
    \begin{tabular}{|c|c|}
     \toprule
     & Audio (e.g. beep)  \\ \midrule
     & Visual (e.g. flashing LED light) \\ \midrule
     & Tactile (e.g. vibration)   \\ \midrule
     & A combination of the above  \\
     \bottomrule
    \end{tabular}%
\end{table}}%
\end{enumerate}